\documentclass[twocolumn,english,aps,arxiv,superscriptaddress]{revtex4-1}%
\usepackage{grffile}
\usepackage[latin9]{inputenc}
\usepackage{color}
\usepackage{babel}
\usepackage{graphicx}
\usepackage{epstopdf}
\usepackage{amsmath}
\usepackage{cancel}
\usepackage{multirow}
\usepackage{ulem}
\usepackage[unicode=true, bookmarks=false, breaklinks=false,pdfborder={0 0
1},colorlinks=false]{hyperref}
\usepackage{breakurl}
\usepackage{amsfonts}
\usepackage{amssymb}%
\providecommand{\U}[1]{\protect\rule{.1in}{.1in}}
\newcommand{\bet}{\begin{table}[hbt]\centering}
\PassOptionsToPackage{normalem}{ulem}
\begin{document}
\title{Tuning the magnetism and 
band topology through antisite defects in Sb doped MnBi$_4$Te$_7$}
\author{Chaowei Hu}
\email{These authors contribute equally.}
\affiliation{Department of Physics and Astronomy and California NanoSystems Institute,
University of California, Los Angeles, CA 90095, USA}
\author{Shang-Wei Lien}
\email{These authors contribute equally.}
\affiliation{Department of Physics, National Cheng Kung University, Tainan 701, Taiwan}
\author{Erxi Feng}
\affiliation{Neutron Scattering Division, Oak Ridge National Laboratory, Oak Ridge,
Tennessee 37831, USA}
\author{Scott Mackey}
\affiliation{Department of Physics and Astronomy and California NanoSystems Institute,
University of California, Los Angeles, CA 90095, USA}
\author{Hung-Ju Tien}
\affiliation{Department of Physics, National Cheng Kung University, Tainan 701, Taiwan}
\author{Igor I. Mazin}
\affiliation{Department of Physics and Astronomy, and Center of Quantum Science and
Engineering, George Mason University, Fairfax, VA 22030, USA }
\author{Huibo Cao}
\affiliation{Neutron Scattering Division, Oak Ridge National Laboratory, Oak Ridge,
Tennessee 37831, USA}
\author{Tay-Rong Chang}
\email{Corresponding author: u32trc00@phys.ncku.edu.tw}
\affiliation{Department of Physics, National Cheng Kung University, Tainan 701, Taiwan}
\affiliation{Center for Quantum Frontiers of Research and Technology (QFort), Tainan 701, Taiwan}
\affiliation{Physics Division, National Center for Theoretical Sciences, National Taiwan University, Taipei, Taiwan}
\author{Ni Ni}
\email{Corresponding author: nini@physics.ucla.edu}
\affiliation{Department of Physics and Astronomy and California NanoSystems Institute,
University of California, Los Angeles, CA 90095, USA}

\begin{abstract}
The fine control of magnetism and electronic structure in magnetic topological insulator is crucial in order to realize the various novel magnetic topological states including axion insulators, magnetic Weyl semimetals and Chern insulators etc. Through crystal growth, transport, thermodynamic, neutron diffraction measurements, we show that under Sb doping the newly-discovered intrinsic antiferromagnetic (AFM) topological insulator MnBi$_{4}$Te$_{7}$ evolve from AFM to ferromagnetic (FM) and then ferrimagnetic. We attribute this to the formation of Mn$_{\rm{(Bi, Sb)}}$ antisites upon doping, which results in additional Mn sublattices that modify the delicate interlayer magnetic interactions and cause the dominant Mn sublattice to go from AFM to FM. We further investigate the effect of antisites on the band topology using the first-principles calculations. Without considering antisites, the series evolves from AFM topological insulator ($x=0$) to FM axion insulators. In the exaggerated case of 16.7\% of periodic antisites, the band topology is modified and type-I magnetic Weyl semimetal phase can be realized at intermediate dopings. Therefore, this doping series provides a fruitful platform with continuously tunable magnetism and topology for investigating emergent phenomena, including quantum anomalous Hall effect, Fermi arc states, \textit{etc}.

\end{abstract}

\date{\today}
\maketitle



\section{Introduction}
Magnetic topological material provides a great platform in discovering new topological states, such as axion insulators, magnetic Weyl semimetals and Chern insulators \cite{tokura2019magnetic}. Emergent phenomena including quantum anomalous Hall (QAH) effect and quantized magnetoelectric effect have been proposed or observed in these phases, offering unprecedented technological opportunities to low-energy-consumption devices, quantum metrology and quantum computing \cite{he2018topological,liu2016quantum,wang2015quantized}.
Recently, MnBi$_{2n}$Te$_{3n+1}$ (MBT) series are shown to be intrinsic magnetic topological insulators (MTIs), where QAH effect was observed \cite{lee2013crystal, rienks2019large,zhang2019topological,li2019intrinsic, otrokov2019prediction, aliev2019novel,147, 1813, deng2020high-temperature,gong2019experimental,lee2019spin, yan2019crystal,zeugner2019chemical,otrokov2019unique,zhang2019topological,chen2019intrinsic,wu2019natural,hao2019gapless,deng2020quantum,liu2020robust,chen2019topological,li2019dirac,li2020competing,ding2020crystal,shi2019magnetic,tian2019magnetic,yan2020type,gordon2019strongly,hu2020universal,xu2019persistent,jo2020intrinsic,tian2019magnetic,klimovskikh2020tunable}. They are made of alternating $n-1$ quintuple-layered (QL) blocks of [Bi$_2$Te$_3$] and one septuple layer (SL) of [MnBi$_2$Te$_4$]. The great structural tunability and natural heterostructural nature of MBT have made it a rare and unique platform to study the interplay among magnetism, band topology, electron correlations and crystal structure. By reducing the interlayer coupling with increasing $n$ and thus the interlayer Mn-Mn distance, MBT evolves from a Z$_2$ A-type AFM TI with saturation fields of 8 T in MnBi$_2$Te$_4$ ($n=1$) \cite{otrokov2019prediction,yan2019crystal,lee2019spin}, 0.22 T in MnBi$_4$Te$_7$ ($n=2$) \cite{147,wu2019natural}, 0.18 T in MnBi$_6$Te$_{10}$ ($n=3$) \cite{shi2019magnetic,tian2019magnetic,yan2020type,gordon2019strongly}, to a ferromagnetic (FM) axion insulator in MnBi$_8$Te$_{13}$ ($n=4$) \cite{1813}. Continuous tuning of magnetism and band topology was achieved in Sb doped MnBi$_2$Te$_4$ \cite{chen2019intrinsic,
yan2019evolution,ko2020realizing}. This series goes from the AFM TI state in MnBi$_{2}$Te$_{4}$ to
likely trivial ferrimagnetic state in MnSb$_{2}$Te$_{4}$ where an additional Mn
sublattice arises from the Mn$_{\rm{(Bi, Sb)}}$ antisite defects \cite{yan2019evolution, murakami2019realization,liu2021site,du2021tuning}. Moreover, angle-resolved photoemission spectroscopy (ARPES) measurements reveal that even 7.5\% of Sb doping induces sizable surface gap opening in MnBi$_2$Te$_4$ \cite{ko2020realizing}. The delicate energy-scale competition 
was further demonstrated in MnSb$_2$Te$_4$ \cite{liu2021site}. While both Mn sublattices always couple antiferromagnetically, in samples
with slightly lower (higher) Mn$_{\rm{Sb}}$ antisite defects, each Mn sublattice orders antiferromagnetically (ferromagnetic ally)\cite{liu2021site}.

Compared to MnBi$_2$Te$_4$, the
AFM and FM ground states can be closer in energy in high $n$ MBT due to the larger Mn-Mn interlayer distance.
Moreover, additional Mn sublattices will arise from the antisites in QLs upon doping \cite{du2021tuning}. Therefore, the high $n$ MBT will be more sensitive to perturbation. Indeed, 30\% of Sb doping makes
MnBi$_{6}$Te$_{10}$ FM \cite{wu2020toward}. However, due to the lack of
intermediate doping levels, it is unclear how the magnetism and band topology
develop in the high $n$ members. Here we report the study of the
effect of Sb doping on MnBi$_{4}$Te$_{7}$ and reveal the important role of the
Mn$_{\rm{(Bi, Sb)}}$ antisites upon Sb doping in governing the magnetism and
band topology.

\section{Methods}
Single crystals were grown using the flux method\cite{yan2019crystal,147}. Elemental forms of Mn, Bi, Sb and Te are mixed at the ratio of MnTe:(Bi$_{1-x}$Sb$_x$)$_2$Te$_3$=15:85 in an alumina crucible sealed in a quartz ampule under $1/3$ atm of argon. The quartz ampule is quickly heated to 900$^{\circ}$C and stay for 5 hrs, followed by a quick cooling to ten degrees above the targeted spin-out temperature. Then it is slowly cooled to the spin-out temperature over three days and stays for another three days before the spin-out. Since each of the doped sample has a very narrow but different growth temperature window, similar trial-and-error growth strategy to the one for MnBi$_8$Te$_{13}$ was used to determine the spin-out temperature \cite{1813} here. Mn doped Bi$_2$Te$_3$ (023 phase) can intergrow with the targeted phase and may contaminate the intrinsic properties. Therefore, extra care was paid to select the best pieces for the study. First, x-ray diffraction at low angles for both the top and bottom (00l) surfaces were measured to select the ones without the 023 phase on the surface. Then, a part of the selected piece was ground for the powder x-ray diffraction (PXRD) to further screen the samples without the 023 impurities. Finally, all measurements were performed on the same piece. We found no significant piece-to-piece variation in magnetic properties within each batch. 

Chemical analysis was obtained by the wavelength-dispersive spectroscopy (WDS) measurements performed on a JEOL JXA-8200 Superprobe. Thermodynamic measurement was performed using the QD Magnetic Properties Measurement System (QD MPMS3). Electric transport measurement was performed using the standard 6-probe configuration using the Quantum Design Dynacool Physical Properties Measurement System (QD Dynacool PPMS). To eliminate unwanted contributions from mixed transport channels of the magnetotransport data, data were collected while sweeping the magnetic field from -9 T to 9 T. The data were then symmetrized to obtain $\rho_{xx}(B)$ using $\rho_{xx}(B)=\frac{\rho_{xx}(B)+\rho_{xx}(-B)}{2}$ and antisymmetrized to get $\rho_{yx}(B)$ using $\rho_{yx}(B)=\frac{\rho_{yx}(B)-\rho_{yx}(-B)}{2}$. The sign of $\rho_{yx}$ is chosen so that hole carriers lead to positive $\rho_{yx}$. 

Single-crystal neutron diffraction was performed for the $x=$0.76 sample at 5 K and 50 K at 0 T on the HB-3A DEMAND single-crystal neutron diffractometer located at Oak Ridge National Laboratory \cite{chakoumakos2011four}. Neutron wavelength of 1.551 $\text{\AA}$ was selected by a bent perfect Si-220 monochromator. The nuclear and magnetic structures were subsequently refined with the FULLPROF SUITE software \cite{rodriguez1993recent}.

We computed the electronic structures using the projector augmented wave method \cite{blochl1994projector,kresse1999ultrasoft} as implemented in the VASP package \cite{kresse1996efficiency} within the generalized gradient approximation schemes (GGA) \cite{perdew1996generalized} and GGA plus Hubbard U (GGA + U) \cite{dudarev1998electronic} scheme. On-site $U =$ 5.0 eV was used for Mn d orbitals. A 11$\times$11$\times$5 MonkhorstPack $k$-point mesh was used in the computations. The spin-orbit coupling effects were included in calculations. The experimental lattice parameters were used. The atomic positions were relaxed until the residual forces were less than 0.01 eV/$\AA$. We used Mn $d$ orbitals, Bi $p$ orbitals, Sb $p$ orbitals, and Te $p$ orbitals to construct Wannier functions, without performing the procedure for maximizing localization \cite{marzari1997maximally}.

\begin{figure}
\centering
\includegraphics[width=3.4in]{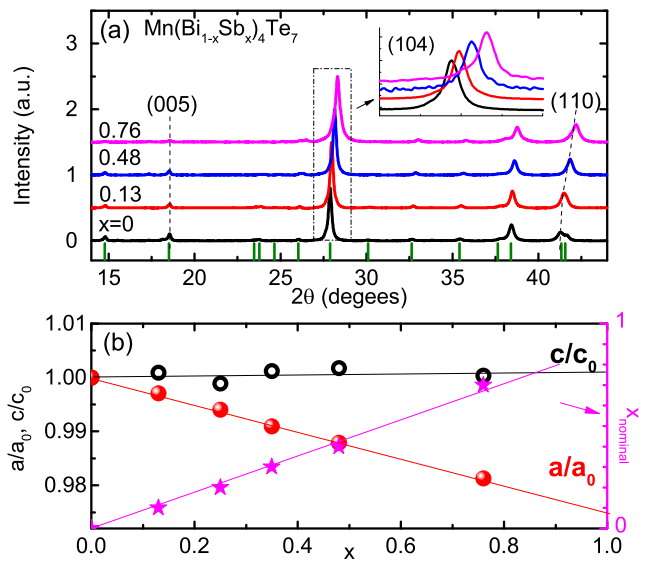} \caption{(a) PXRD of Mn(Bi$_{1-x}$Sb$_x$)$_4$Te$_7$ for the pieces for which the data in the paper were collected.  The peak positions of the 147 phase are marked. Inset: the zoom-in plot of the (104) PXRD peaks. (b) The doping-dependent relative lattice parameters $a/a_0$, $c/c_0$ and nominal concentration $x_{\text{nominal}}$ used in the growth. $a_0$ and $c_0$ are the lattice parameters for MnBi$_4$Te$_7$.}
\label{fig:xrd-lp}
\end{figure}

\section{Results}

\begin{figure*}[ptb]
\centering
\includegraphics[width=7in]{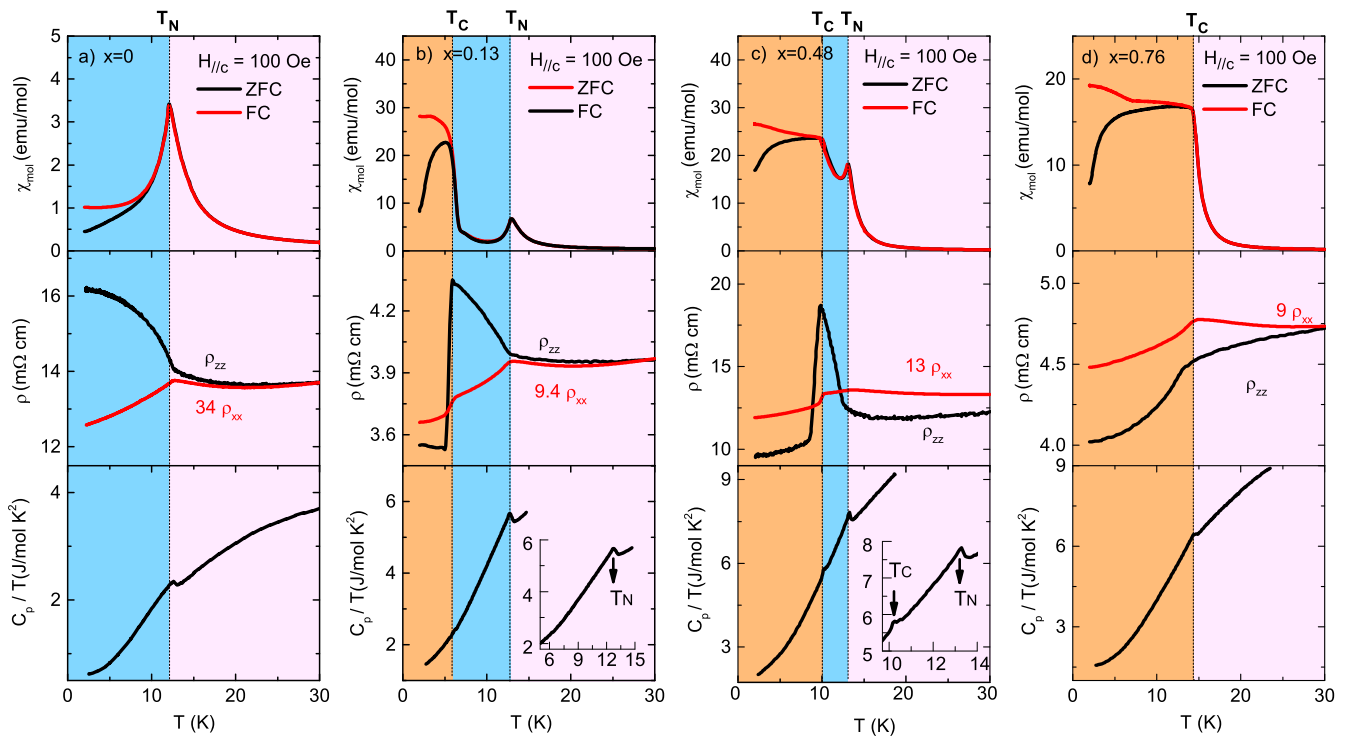} \caption{The evolution of magnetism with temperature in Mn(Bi$_{1-x}
$Sb$_{x}$)$_{4}$Te$_{7}$ from $x=0$ to $x=$0.76: (a)-(d) Top panel: $\chi^{c}(T)$, the
temperature-dependent zero-field-cool (ZFC) and field-cool (FC) magnetic
susceptibility under 0.01 T with $H\|c$. Middle panel: $\rho_{xx}(T)$ and $\rho_{zz}(T)$, the
temperature-dependent electrical resistivity with the current
along the $ab$ plane and the $c$ axis. Bottom panel: $C_{p}(T)$, the temperature-dependent specific heat; and the zoom-in of $C_{p}(T)$ near transitions (inset). Note: the
resistivity curves of $x=0$ are from Ref. \cite{147}.}%
\label{fig:transport-T}%
\end{figure*}

Both the PXRD and chemical analysis via WDS indicate that Sb was successfully doped into MnBi$_4$Te$_7$ and the results are summarized in Fig. 1 and Table I. Figure 1(a) shows the PXRD patterns for various doping levels. All peaks can be indexed by the 147 phase. If there is 023 phase impurity, a clear hump will appear at the left shoulder of the (104) peak. As shown in the inset of Fig. 1(a), the 023 phase is indiscernible or less than 5\% if there is. With Sb doping, the (005) peak roughly stays at the same angle while the (104) peak shifts moderately and the (110) peak shifts much to higher angles, indicating distinct in-plane and out-of-plane lattice response to the Sb doping. Figure 1(b) shows the doping-dependent lattice evolution. The lattice parameter $a$ decreases linearly by 2\% up to our highest doping level $x= 0.76$ while the lattice parameter $c$ remains almost the same. This lattice evolution is similar to that in Mn(Bi$_{1-x}$Sb$_x$)$_2$Te$_4$ \cite{yan2019evolution}.

WDS reveals the real doping level of Sb as well as a universal deficiency of Mn in all compounds as seen in Table I. The total Mn concentration is near 0.8 for all but slowly increases with $x$. This is because when Sb substitutes Bi, it also introduces the preferable Mn$_{\rm{Sb}}$ antisites
\cite{murakami2019realization,liu2021site,yan2019evolution,du2021tuning} (high-spin Mn$^{2+}$ ionic radius, 99 pm, is closer to that of Sb$^{3+}$, 90 pm, than to
Bi$^{3+}$, 117 pm\cite{webelements}). Therefore, upon Sb doping, more and more Mn can enter into the Bi/Sb sites. In contrast to the Mn(Bi$_{1-x}$Sb$_x$)$_2$Te$_4$ series where two Mn sublattices exist due to the Mn$_{\rm{Sb}}$ antisite formation, Sb doping in MnBi$_4$Te$_7$ gives rise to a more complex antisite chemistry, where three Mn sublattices appear. We
denote the Mn atoms occupying the Mn site as Mn1 sublattice, the Mn atoms on
the Bi site within SLs as Mn2 sublattice, and the Mn atoms on the Bi site in
QLs as Mn3 sublattice, see Fig. 1 in \cite{147}. As we will show later, while the Mn2 and Mn3 antisites are not as concentrated as Mn1, they do make a big impact on the overall magnetism and band topology.

\begin{figure*}[ptb]
\centering
\includegraphics[width=7in]{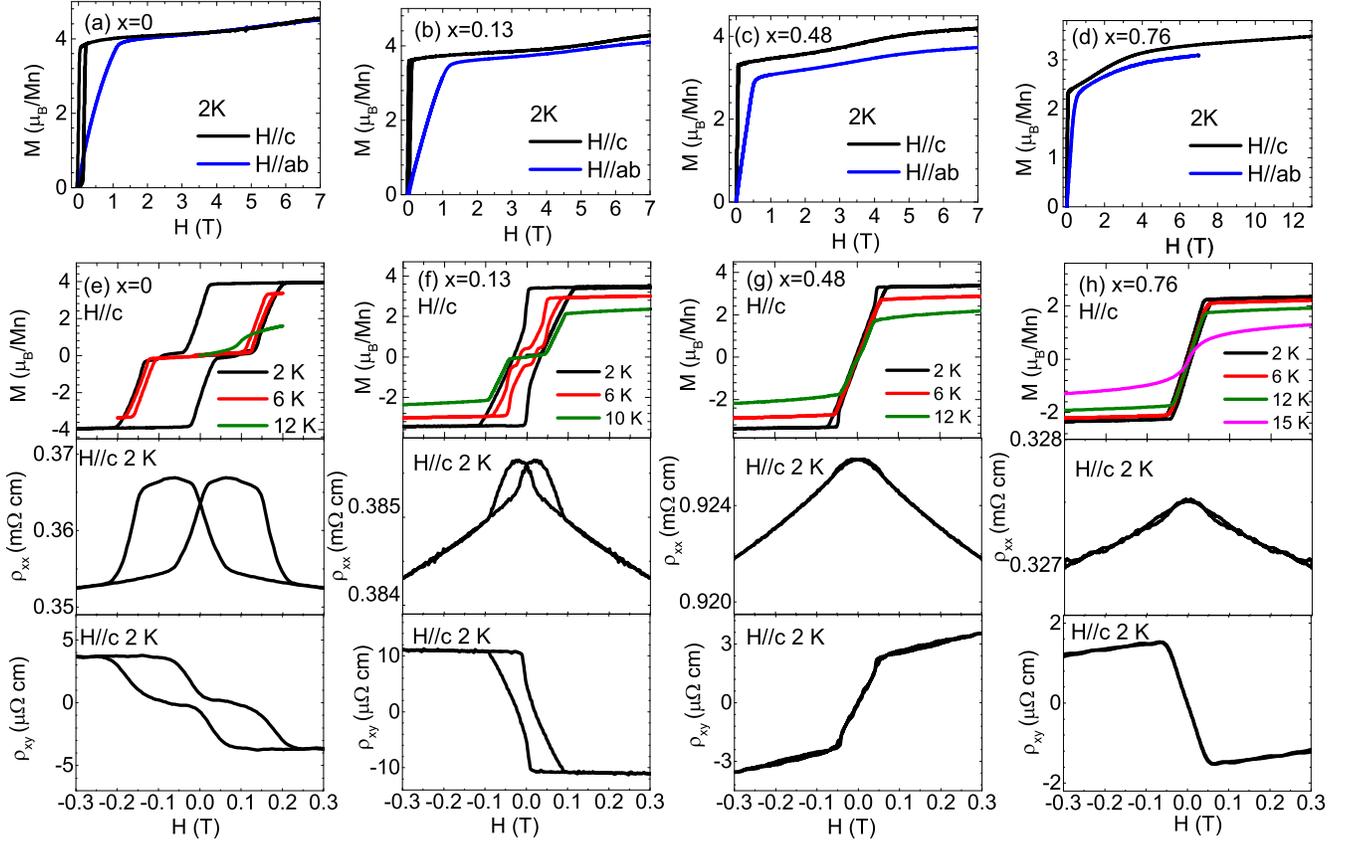} \caption{ The evolution of magnetism with field in Mn(Bi$_{1-x}
$Sb$_{x}$)$_{4}$Te$_{7}$ from $x=0$ to $x=$0.76: (a)-(d) The isothermal
magnetization $M(H)$ at 2 K with $H\|c$ and $H\|ab$ up to 7 T. The unit is chosen as $\mu_{B}$/Mn where the Mn
concentrations via WDS data are used. (e)-(g): The hysteresis of $M(H)$ (top), $\rho_{xx}(H)$ (middle) and $\rho_{xy}(H)$ (bottom)
with $H\|c$ at 2 K (unless noted otherwise). At 2K, the hysteresis of the $M(H)$ curve goes from AFM type $(x=0)$ to FM type $(x=0.76)$. }%
\label{fig:transport-H}%
\end{figure*}

\subsection{Evolution of magnetism}

The evolution of magnetic structure in the series can be well traced in the temperature-dependent susceptibility with $H\|c$ ($\chi^{c}(T)$), the temperature-dependent resistivity with $I\|ab$ ($\rho_{xx}(T)$) and $I\|c$ ($\rho_{zz}(T)$), as well as the temperature-dependent specific heat ($C_{p}(T)$) data in Fig. 1. For $x=0$, at
$T_{N}=12.7$ K, the sharp cusp in $\chi^{c}(T)$ and the kink in $C_{p}(T)$
signal the paramagnetic (PM) to AFM phase transition. Furthermore, below
$T_{N}$, the sudden increase of $\rho_{zz}(T)$ and drop of $\rho_{xx}(T)$ are
indicative of the gain (loss) of the spin-disorder scattering along the $c$
axis ($ab$ plane). This suggests spins parallel in the $ab$ plane but
anti-parallel along the $c$ axis, consistent with the A-type AFM structure
revealed by neutron data \cite{yan2020type, ding2020crystal}.

For $x=0.13$, in addition to a sharp cusp at $T_{N}=12.8$ K in
$\chi^{c}(T)$ suggesting AFM transition like the parent compound, a second transition occurs at $T_{C}=6.0$ K marked by a sudden
increase and a large bifurcation in the ZFC and FC data of $\chi^{c}(T)$. From
$T_{N}$ to $T_{C}$, $\rho_{zz}(T)$ increases sharply and $\rho_{xx}(T)$ drops,
analogous to that of the $x=0$ sample, while below
$T_{C}$, $\rho_{zz}(T)$ decreases sharply due to the loss of spin scattering like that in the FM MnBi$_{8}$Te$_{13}$ \cite{1813}. Meanwhile, since Mn1 and Mn2 sublattices in the SL plane are strongly AFM coupled to each other through superexchange interaction
\cite{liu2021site}, Mn2 sublattice most likely orders simultaneously and antiferromagnetically with Mn1
at $T_{N}$. Therefore, we argue that from $T_{N}$ to $T_{C}$, Mn1 and Mn2 sublattices each adopts
A-type AFM, while they also AFM coupled to each other. Below $T_{C}$, while Mn1 and Mn2 sublattices are still AFM coupled to each other, they are FM within themselves.
We denote these two magnetic structure as Ferri$^{AFM}$ and Ferri$^{FM}$, as depicted in Fig. 4(a).

For $x=0.48$, the shapes of the $\chi^{c}(T)$, $\rho_{zz}(T)$ and $\rho
_{xx}(T)$ curves are similar to those of the $x=0.13$ compound, manifesting a
Ferri$^{AFM}$ state between $T_{N}=13.3$ K and $T_{C}=10.2$ K and a Ferri$^{FM}$ state below $T_{C}$. However, in sharp contrast to the $x=0.13$ sample where specific heat anomaly only appears at $T_N$, an additional small specific heat anomaly emerges at $T_{C}$ for $x=0.48$, evidencing an entropy release which is not directly originated from the AFM-FM transition of the Mn1 and Mn2 sublattice at $T_{C}$. Since it is natural to believe that Mn3 concentration is higher in this doping level than that in the $x=0.13$ sample. This additional specific heat release is very likely to be related to the increasing amount and the magnetic state of the Mn3 sublattice.  

For $x\ge 0.58$ (Fig. S1(c)-(d) and Fig. 1(d)), only one phase transition is observed. As a representative, the data for $x=0.76$ is shown in Fig. 1 (d). The $\chi^{c}(T)$, $\rho_{zz}(T)$ and $\rho_{xx}(T)$ are
reminiscent of those of the $x=0.13$ and 0.48 compounds in the Ferri$^{FM}$
state, suggesting Mn1 and Mn2 sublattices order simultaneously at 14.5 K into
the Ferri$^{FM}$ state.  

Figures 3(e)-(h) show the magnetic hysteresis loop, the field-dependent electrical resistivity ($\rho_{xx}(T)$) and Hall resistivity ($\rho_{xy}(T)$) with $H\|c$. A spin flip transition with hysteresis was observed at 0.15 kOe in the $x=0$ sample. The envelopes of the $M(H)$ curves in the Ferri$^{FM}$ ground state are nontrivial. Instead of a standard FM hysteresis loop, a bow-tie-shaped hysteresis loop can be clearly observed at 2 K for the $x=0.13$, 0.37, 0.48 and 0.58 samples. A recent magneto-optic study suggests this bow-tie-shaped hysteresis may be related to the formation of low-field fine-structured isotropic domains and high-field less isotropic sea-urchin-shaped domains \cite{MBTrelaxation}. We also
note that the multi-step feature of the $M(H)$ curve at
6 K for the $x = 0.13$ compound is reminiscent of that
of the MnBi$_6$Te$_{10}$ compound \cite{MBTrelaxation}. This may suggest the existence of small amount of FM domains in the AFM state of MnBi$_6$Te$_{10}$. Therefore, close energy scales of FM and AFM is universal
in the $n \geq 2$ MBT members. This explains the controversies on the magnetic
ground states of MnBi$_4$Te$_7$ obtained from different growth methods, while the first-principle calculations suggest an AFM ground state \cite{147,wu2019natural}. This is all because
slight site defects are enough to tune the exchange energy
to surrender one and boost the other.

Magnetism and charge carriers are strongly coupled. At $x=0$, a sharp drop in $\rho_{xx} (T)$ can be observed at the spin-flip field around 0.15 T due to the loss of spin-disorder scattering. On the other hand, $\rho_{xx} (T)$ shows subtle decrease for the $x=0.13$, 0.48 and 0.76 samples, consistent with their Ferri$^{FM}$ ground state. In the Hall resistivity panels, anomalous Hall effect arising from the internal magnetization can be seen in all concentrations. Furthermore, a clear sign change of ordinary Hall resistivity appears at $x=$0.48. We then determine the carrier density using $n=\frac{1}{R_H e}$ where the Hall coefficient $R_H$ is the slope of $\rho_{xy}(H)$ at 20 K in Fig. S2. After interpolating the 
carrier density with doping level, an estimated charge neutrality point is found near $x=0.36$, as shown in Fig. S2(b).

\subsection{Defects and the magnetic state of the Mn3 sublattice}

Based on our aforementioned discussions about Fig. 1, we have hypothesized the magnetic structures of the Mn1 and Mn2 sublattices. However, a few important questions remain unclear for this doping series. What are the concentrations of the Mn1, Mn2 and Mn3 sublattices? Does the Mn3 sublattice order? If yes, what is its ordering state? If not, is it glassy or fluctuating? Single-crystal neutron scattering measurements were performed on the $x=0.76$ sample, together with the high-field $M(H)$ data which provides an alternative way to estimate the concentrations of the three Mn sublattices, they shed light on these questions.

\begin{table*}[]
\setlength{\tabcolsep}{6pt}
\renewcommand{\arraystretch}{2}
\caption{Chemical composition of Sb doped Mn-147, the estimation of defects concentration with $m$, $n$, and $l$ representing the Mn occupancy on Mn2 sites, Mn occupancy on Mn3 sites, and Mn occupancy on Mn1 sites (see text). The magneocrystalline anisotropy parameter is estimated from the saturation field of the dominant Mn1 sublattice ($H_{sat}^{ab}$)$_{\rm{Mn1}}$ (see text).}
\label{table:wds}
\begin{tabular}{c|c|c|c|ccc|c}
\hline 
\multirow{4}{*}{WDS}& $x_{\rm{nominal}}$ & Mn : Bi : Sb : Te  & Sb/(Bi+Sb) & $m$ & $n$ & $l$ & SD (meV) \\ \hline 
&0     & 0.78(3) : 4.27(7) : 0 : 7 & 0   & 0.015(5)  & 0.03(1)  & 0.73(3) & 0.069 \\
&0.1    &0.79(2) : 3.70(3) : 0.57(1) : 7 & 0.13(1) & 0.020(5)  & 0.04(1)  & 0.70(3) & 0.069 \\
&0.4    &0.81(4) : 2.19(5) : 2.04(1) : 7 & 0.48(1) & 0.025(5)  & 0.055(10)  & 0.65(3) & 0.029 \\
&0.7     &0.82(1) : 1.00(3) : 3.08(2) : 7 & 0.76(1) & 0.060(5)  & 0.07(1)  & 0.56(3) &  0.017\\\hline
neutron& 0.7     & 0.88(3): 1.15(6) : 2.96(6) : 7 & 0.72(2) & 0.06(1)  & 0.07  & 0.62(1) &  0.017 \\ \hline
\end{tabular}
\end{table*}

The preliminary refinement of the single-crystal neutron data with all fitting parameters free indeed indicates a Ferri$^{FM}$ state of the Mn1 and Mn2 sublattice at 5 K, consistent with Fig. 4(a). The refinement further suggests the moment contributed from Mn3 atoms is 0.13(10)$\mu_B$/formula-unit. This small moment either implies negligible Mn3 concentration, or Mn3 sublattice is glassy/fluctuating. To differentiate these two scenarios, let us focus on the high-field $M(H)$ data with $H\|c$ shown in Figs. 3(a)-(d).
Above a sharp increase caused by the Mn1+Mn2 complex below 0.2 T, a universal, subtle, but clear medium-field magnetization increase appears around 1-5 T in the whole doping series. Because the coupling between Mn1 and Mn3 is weaker than that between Mn1 and Mn2, Mn3 should polarize at a lower field than Mn2. Therefore, this medium-field magnetization increase should arise from the polarization of the Mn3 sublattice since the Mn2 sublattice will flip at a much higher fields. Indeed, for $x=0.76$, $M(H)$ is only 3.4 $\mu_{B}$/Mn at 13 T. Despite reaching a plateau, it is still much smaller than 5 $\mu_{B}$/Mn, the theoretical value for Mn$^{2+}$. This strongly indicates the flipping of Mn2 will happen at a field higher than 13 T. As a reference, in MnBi$_2$Te$_4$, the flipping process of the Mn2 sublattice starts at 20 T and ends at 60 T \cite{jiaqiang_highfield}.

Such a field-dependent magnetic structure evolution is depicted in Fig. 4(b), which allows us to separate the contribution of magnetization from each Mn sublattices in the $M(H)$ curve with $H\| c$ (Figs. 3(a)-(d)). Right before the sizable polarization of Mn3 sublattice, the sample is in the state I depicted in Fig. 4(b) and we denote the moment to be $\alpha$. Then with increasing field, more and more Mn3 atoms are polarized into the state II as pictured in Fig. 4 (b), leading to a plateau in $M(H)$ with the moment of $\beta$. With even higher fields, Mn2 will be polarized and finally all three Mn sublattices stay in state III as drawn in Fig. 4 (b), resulting in a moment of $\gamma$. Therefore, the difference in $\alpha$ and $\beta$ can tell us the total moment contributed by Mn3 atoms under field. For $x=0.76$, this value is $\sim$ 0.8(1) $\mu_B$/Mn, suggesting the Mn3 spins contribute significantly to the ordered moment under fields and thus Mn3 concentration is not negligible. Together with the neutron scattering data which reveal all Mn3 atoms only contribute 0.13(10)$\mu_B$/formula-unit at zero field, we conclude that Mn3 atoms are in the glassy/fluctuating state at 5 K. Furthermore, considering that Mn3 atoms are mostly polarized above 1 T while the recent AC susceptibility measurements of the $x=0.76$ sample only show relaxation behavior below 500 Oe\cite{hu2021unusual}, it is likely that at 5 K the Mn3 atoms are not in the glassy state, but rather fluctuating in the paramagnetic state. Future site-sensitive nuclear magnetic resonance measurements can help clarify the magnetic state of Mn3 sublattice. 

\begin{figure}
\centering
\includegraphics[width=3.4in]{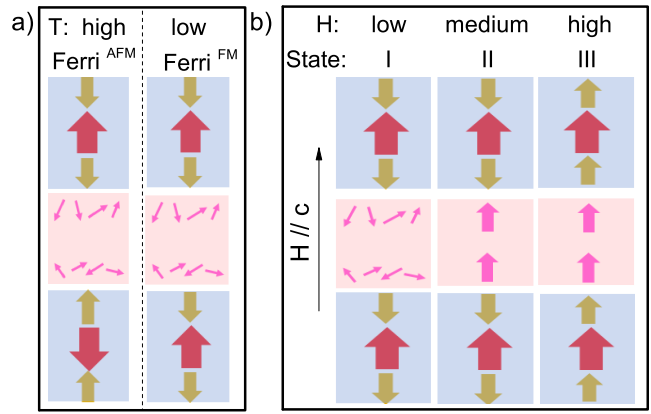} \caption{a) The depiction of the Ferri$^{AFM}$
and Ferri$^{FM}$ states. Red, olive and pink arrows represent the
Mn1, Mn2 and Mn3 spins. b) Ferri$^{FM}$ state with $H\|c$ at low, medium and high field: the depiction of the process of the polarization of Mn3 and the spin-flipping of Mn2.}
\label{fig:summary}
\end{figure}

According to the sequential field-induced processes illustrated in Fig. 4 (b), we can quantitatively estimate the concentration of the three Mn sublattices based on the $M(H)$ data. If we set the amount of the Mn2 and Mn3 antisites and the Mn occupancy on the Mn1 site to be $m$, $n$ and $l$ respectively, then the concentration of the Mn1, Mn2 and Mn3 sublattices is $1-l$, $2m$ and $2n$ with $2m+2n+(1-l)=Mn^{\rm{WDS}}$ where Mn$^{\rm{WDS}}$ is the total Mn concentration determined by the WDS measurement. Assuming the ordered-moment/Mn for Mn1, Mn2 and Mn3 is the same, we can readily write down a set of equations based on Fig. 4(b):
\begin{align}
  & (l-2m)/\rm{Mn^{\rm{WDS}}}=\alpha/\gamma \\
  & (l-2m+2n)\rm{/Mn^{\rm{WDS}}}=\beta/\gamma \\
  & (2m+2n+l)/\rm{Mn^{\rm{WDS}}}=1
\end{align}

For the $x=0.76$ sample, Mn$^{\rm{WDS}}=0.82$. We take the moment at 1 T as $\alpha$ where the subtle slope change in $M(H)$ suggests the onset of the polarization process of the Mn3 sublattice. We then take the moment at 13 T as $\beta$ and assume $\gamma=5$ $\mu_B$/Mn. Using these values, we estimate $m$, $n$ and $l$ to be 0.060(5), 0.07(1) and 0.56(3). For the other dopings, since $M(H)$ was only measured up to 7 T, we set $\beta$ to be the moment at 7 T plus 0.1$\mu_B$/Mn. The obtained $m$, $n$ and $l$ are summarized in Table I.

\begin{figure*}[ptb]
\centering
\includegraphics[width=7in]{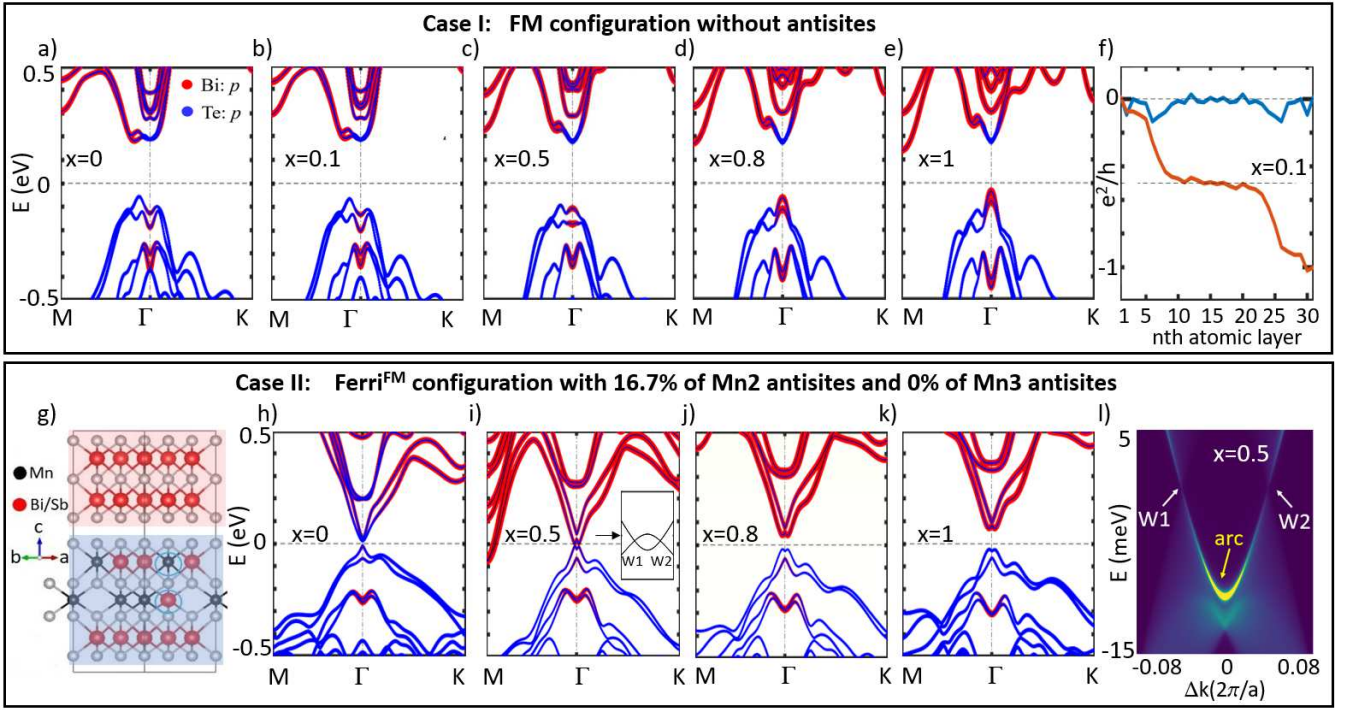} \caption{(a)-(f) DFT calculations in
the defect-free and FM configuration case: (a)-(e) Bulk band structures. The
red and blue dots indicate (Bi,Sb)-$p$ and Te-$p$ orbitals, respectively. (f)
Layer-resolved AHC for $x=0.1$ sample. Partial AHC of each atomic layer (blue
line); Integral of the partial AHC (red line). (g)-(l) DFT calculations in the
16.7\% of periodic Mn2 antisites and Ferri$^{FM}$ configuration case: (g) The depiction of the
structure model used. (h)-(k) Bulk band structures. (l) Surface band structure
($x=0.5$) along the momentum space cut that goes through a direct pair of
Weyl nodes W1 and W2. }%
\label{fig:topology}%
\end{figure*}

Keeping the afore discussed defect estimation in mind, next let us switch gear back to the neutron data. Quantitatively, since this is a doped system with multiple types of defects, we have to make several assumptions for the refinement. Firstly, for the magnetic reflection data taken at 5 K, we fix the ordered moment of Mn3 as 0 and the concentration of Mn3 as 0.07; we then assume a fixed moment of 4.6 $\mu_B$/Mn for Mn1 and Mn2, the same value as that of the parent MnBi$_4$Te$_7$ obtained from neutron scattering \cite{ding2021neutron}. By these restrictions, Mn1 and Mn2 occupancy are refined. Secondly, the Mn1 and Mn2 occupancies are then used for a more comprehensive structural refinement for the scattering data at 50 K to better determine the atomic coordinates and the Sb and Bi level. Such information is then fed back to the magnetic refinement of the 5 K data. A recursive process are repeated until all values converge. Eventually, the refinement with 27 reflections at 5 K yielded R$_{\rm{F}}$=3.03\% and $\chi^2$ = 1.99; the refinement with 118 reflections at 50 K yielded R$_{\rm{F}}$=2.16\% and $\chi^2$ = 1.46. The refinement result is summarized in Table II and Table I. Our refinement unambiguously shows site mixing. Opposite magnetic moments are observed in Mn1 and Mn2 sites, suggesting the Ferri$^{\rm{FM}}$ state. $m$ is 0.06(1), $l$ is 0.62(1) and the total Mn concentration is 0.88(3). Considering that the Mn3 concentration is fixed with no error in the neutron refinement which will lead to a smaller error bar in the total Mn concentration, these defect concentrations are consistent with the ones obtained from the $M(H)$ and WDS data. In addition, Sb atoms are found to be inhomogeneously doped in each site. The Sb has an overall higher concentration in the SLs than in QLs. In the Mn1 site, no Bi is found at all.

\begin{table}[]
\setlength{\tabcolsep}{3pt}
\renewcommand{\arraystretch}{1.5}
\caption{Refined structural parameters for the $x=$0.76 sample based on the single crystal neutron diffraction data measured at 50 K and 5 K.}
\centering
\begin{tabular}{cccccc}
\hline
Name & $x$     & $y$     & $z$          & occ.     & Moment at 5 K         \\
\hline\hline
Mn1  & 0     & 0     & 0.5         & 0.623(4) & 4.6 $\mu_B$/Mn           \\
Bi1  & 0     & 0     & 0.5        & 0        &               \\
Sb1  & 0     & 0     & 0.5        & 0.377(4) &               \\\hline
Bi2  & 0.333 & 0.667 & 0.342(1)   & 0.21(2)  &               \\
Sb2  & 0.333 & 0.667 & 0.342(1)      & 0.73(2)  &               \\
Mn2  & 0.333 & 0.667 & 0.342(1)     & 0.06(1)  & -4.6 $\mu_B$/Mn          \\\hline
Bi3  & 0.333 & 0.667 & 0.0842(1)   & 0.37(2)  &               \\
Sb3  & 0.333 & 0.667 & 0.0842(1)   & 0.56(2)  &               \\
Mn3  & 0.333 & 0.667 & 0.0842(1)   & 0.07  & 0             \\\hline
Te1  & 0     & 0     & 0          & 0.083    &               \\
Te2  & 0.333 & 0.667 & 0.155(1)   & 0.167    &               \\
Te3  & 0     & 0     & 0.273(1)    & 0.167    &               \\
Te4  & 0.333 & 0.667 & 0.430(1)   & 0.167    &              \\\hline
\end{tabular}
\end{table}

\subsection{Band topology} 
To investigate the evolution of band topology for this series, DFT calculations are performed. For the $x=0$ compound, the DFT calculations are made in the A-type AFM configuration. The presence of the band inversions in Fig. S3 (a) hints toward a topological phase. To reveal the topological ground state of this system, we compute the topological invariant $Z_2$ index. In general case, $Z_2$ invariant is ill-defined in a time-reversal symmetry breaking system. Nonetheless, $Z_2$ can be expanded to classify topological states in magnetic system that possesses a specific magnetic configuration, for example, an A-type AFM state, in which the combinatory symmetry $S=\theta T_{1/2}$ is existed, where $\theta$ is time-reversal operator and $T_{1/2}$ is the half translational symmetry along the c axis of the AFM primitive cell \cite{mong2010antiferromagnetic,fang2013topological,gui2019new}. The trajectory of Wannier charge centers (WCCs) is an open curve traversing the whole Brillouin zone (BZ) in the k$_z$ = 0 plane via the $Z_2$ invariant by the Wilson loop method (Fig. S3(b)), indicating $Z_2$ = 1. Thus, MnBi$_4$Te$_7$ is an antiferromagnetic topological insulator (AFM-TI), similar to the previous report\cite{147}.

For the doped compounds with FM ground states, we discuss two limiting cases. In case I, no defect is considered and FM configuration is used. In case II, large amount of antisites are studied and the Ferri$^{FM}$ order is used. For the sake of feasibility, in case II, we assume only Mn1 and Mn2 sublattices exist and 16.7\% of Bi/Sb atoms in the SLs exchange with the Mn atoms on the same layer to form Mn2 antisites. The value is close to the antisite concentration in MnSb$_2$Te$_4$ \cite{liu2021site}, but much larger than that in our samples. Hence this will give exaggerated effect of antisites.

\subsubsection*{\textit{Case I: the defect-free scenario}}

We first construct a tight-binding Hamiltonian for both FM MnBi$_4$Te$_7$ and MnSb$_4$Te$_7$ using our experimental lattice parameters in Fig. 1. Then the electronic structures of the doped compounds are calculated by a linear interpolation of tight-binding model matrix elements of the Hamiltonians. This approach was successfully applied to investigate the evolution of band topology in BiTlSe$_{1-x}$S$_{x}$ TI and Mo$_{x}$W$_{1-x}$Te$_{2}$ Weyl semimetal \cite{xu2011topological,chang2016prediction}. By that, the calculated band structure without defects are shown in Figs. 5(a)-(e). 

For all doping levels, a sizable band gap of 200--300 meV opens, with a band inversion between the (Bi, Sb)-$p$ and Te-$p$ states in the vicinity of $E_{F}$. Our topological invariant calculations show the Chern numbers to be zero both in the $k_z$ = 0 and $k_z=\pi$ planes. Next, we compute the parity-based higher-order $Z_4$ invariant, which is defined by $Z_4=\sum_{i=1}^8 \sum_{n=1}^{n=occ}[(1+\xi_n (\gamma_i)]\text{2 }mod\text{ 4}$, where $\xi_n (\gamma_i)$ is the parity eigenvalue (+1 or -1) of the n-th band at the i-th time reversal invariant point $\Gamma_i$ and n = occ is the number of occupied bands\cite{turner2012quantized}. The $Z_4$ invariant is well defined for an inversion symmetric system, even in the absence of time reversal symmetry. The odd values of $Z_4$ ($Z_4$=1,3), indicate a Weyl semimetal phase, while $Z_4$ = 2 corresponds to an insulator phase with a quantized topological magnetoelectric effect (axion coupling $\theta = \pi$) \cite{1813,wieder2018axion}. Our calculation shows the $Z_4$ invariant of Mn(Bi$_{1-x}$Sb$_x$)$_4$Te$_7$ with FM configuration to be 2 for all $x$, which suggests a 3D FM axion insulator phase. 

To show the novel physics, taking the $x=0.1$ compound as an
example, we further investigate the anomalous Hall conductivity (AHC) in the 2D limit (Fig. S4) \cite{supplement}.
For a 31 atomic layer symmetric slab, vacuum-[SL-QL-SL-QL-SL]-vacuum, when the
$E_{F}$ is gated inside the energy band gap, the layer-resolved
AHC calculation (the blue line in Fig. 2 (f))
shows that the AHC mainly comes from the atomic layers on the top and bottom
surfaces (about one SL and half QL), and the intensity rapidly decreases to
approximately zero in the middle region of the device. As a result, each
surface ($\lesssim$ 10 atomic layers) contributes (the red line in Fig. 2(f))
$-0.5$ $e^{2}/h$ to AHC, and gives $-1$ $e^{2}/h$ for the whole slab. The
half-integer quantized plateau in the middle indicates that the axion coupling
strength equals the quantization value of $\pi$ in this device, further
confirming the $x=0.1$ compound as a FM axion
insulator\cite{varnava2018surfaces}.

\subsubsection*{\textit{Case II: periodic Mn2-antisite scenario}}

To include the antisite effect in case II, we perform the supercell calculation. First, we construct a $\sqrt{3}\times\sqrt{3}$ supercell of MnBi$_4$Te$_7$. In this model, one atomic Mn layer contain three Mn atoms (Mn1 sublattice). Then we exchange one of Mn atom and one of Bi atom within the SL (Mn2 sublattice), resulting a chemical formula of (Mn$_{0.67}$ Bi$_{0.33}$) (Bi$_{0.833}$ Mn$_{0.167}$)$_2$Te$_4\cdot$Bi$_2$Te$_3$ as shown in Fig. 5(g). By this, we calculate the band structure of MnBi$_4$Te$_7$ with 16.7\% of the antisite disorder. We then perform similar procedure to calculate the band structure of (Mn$_{0.67}$ Sb$_{0.33}$) (Sb$_{0.833}$ Mn$_{0.167}$)$_2$Te$_4\cdot$Sb$_2$Te$_3$. Finally, electronic structure of the doped compounds are calculated by a linear interpolation of tight-binding model matrix elements of the Bi and Sb versions. 

Our calculation indeed shows that in this defect limit, the magnetic configuration of Mn1 and Mn2 is FM individually while Mn1 and Mn2 are coupled antiferromagnetically, forming the Ferri$^{FM}$ ground state. In the Ferri$^{FM}$ state, the bulk band gap of MnBi$_{4}$Te$_{7}$ is greatly reduced to about 20 meV. Meanwhile, the character of the band inversion also alters with doping, as shown in Figs. 5(h)-(k). Consequently, in the intermediate doping region at $x=0.5$ for example, one can see a small energy gap about 10 meV near $\Gamma$ point (inset of Fig. 5 (i)). This tiny gap implies the existence of Weyl semimetal state. Since the gapless Weyl points do not guarantee to locate on the high symmetry point or the high symmetry line, to confirm the chirality of the Weyl nodes, we calculate the chiral charge based on the Wilson loop method. The associated chiral charge for W1(W2) is calculated to be $-1(+1)$ based on the Wilson loop
method (Fig. S5) \cite{supplement}, indicating that they carry opposite chirality and do
form a pair of Weyl nodes. Furthermore, a topological Fermi arc state, the
characteristics of Weyl semimetal, appears in the (100) surface states and
terminates directly at the projected Weyl nodes (Fig. 5(l)). This further
supports the picture of ferrimagnetic type-I Weyl semimetal with only two Weyl nodes.

\section{Discussion}
As shown in Fig. 6 (b), Sb doping leads to a decrease in the Mn1 concentration, meanwhile, it also boosts the number of Mn2 and Mn3 antisites, leading to an overall enhanced Mn$^{\rm{WDS}}$. This defect evolution is responsible for the evolvement of magnetism, as summarized by the temperature-doping phase diagram in Fig. 6(a). 

We argue that the AFM to FM transition of the dominate Mn1 sublattice arises from the delicate competitions between the direct Mn1-Mn1 AFM interlayer interaction and Mn3-assisted Mn1-Mn1 FM interlayer coupling. The magnetic Hamiltonian may be approximated as follows. The Mn1 sublattice itself has strong FM intralayer couplings $J_0$ and weak interlayer AFM interactions $J_1$. The Mn2 and Mn3 sublattices are so sparse that the interactions among themselves are negligible, but they
couple to Mn1 through strong superexchange interactions $J_2$ and $J_3$ due to their proximity. Finally, there is Mn2-Mn3 coupling, which modifies $J_{3}$ into $J_3^{\rm{eff}}$. Note that the sign of $J_3^{\rm{eff}}$ can vary from site to site (for instance, depending on whether the closest Mn neighbor to a Mn is Mn1 or Mn2).

Now, while Mn2 just follows Mn1 and does not
affect the overall Mn1-Mn1 interlayer ordering, Mn3 introduces an effective FM interlayer coupling when they couple to the neighboring Mn1 layers above and below:%
\begin{equation}
H_{\rm{interlayer}}=J_{1}\langle\mathbf{M}_{1}\cdot\mathbf{M}_{1}^{\prime}\rangle+\langle
J_{3}^{\rm{eff}}%
(\mathbf{M}_{1}\cdot\mathbf{M}_{3}+\mathbf{M}_{1}^{\prime}\cdot\mathbf{M}%
_{3})\rangle,
\end{equation}
where $\mathbf{M}_{1}$ and $\mathbf{M}_{1}^{\prime}$ are the local magnetizations of the
two neighboring Mn1 layers, $\mathbf{M}_{3}$ is the local magnetization of the Mn3 bilayer between these Mn1 layers, and the brackets denote the average of the $ab$ plane. The second term equals 0, if Mn1 sublattice is AFM, and $-2\langle|J_{3}^{\rm{eff}}M_{3}M_{1}|\rangle$, if Mn1 sublattice is FM, regardless of the sign of $J_{3}^{\rm{eff}}$; thus it always favors the FM ordering of Mn1 sublattice (that is, the Mn3 mediates FM interlayer coupling of the Mn1 sublattice).  
We can see that if $|c_3J_{3}^{\rm{eff}}M_{3}|>c_1J_{1}M_{1},$ the system orders into the Ferri$^{FM}$ state with the energy $E_1=c_1J_{1}M_{1}^{2}-2c_3|J_{3}^{\rm{eff}}M_{3}M_{1}|,$
otherwise into the Ferri$^{AFM}$ state with
$E_2=-c_1J_{1}M_{1}^{2},$where $c_1$ and $c_3$ are the concentrations of Mn1 and Mn3 sublattice.

This consideration implicitly implies that the susceptibility of the Mn3 subsystem is infinite, which would be true at $T=0$, and if the Mn3-Mn3 interaction can be neglected. Our neutron scattering indicates that Mn3 spins are strongly fluctuating, which suggests that the susceptibility of this subsystem, $\chi_{Mn3}$, is finite. This does not change the conclusions qualitatively; the only change is that if  $\chi_{Mn3}$ is relatively small, the net Mn3 subsystem magnetization $M_3$ becomes proportional to $M_1\chi_{Mn3}(T)$, and thus explicitly $T$-dependent.

With this in mind, we can explain both the temperature and doping evolution of the magnetism. Because of the Mermin-Wagner physics, $M_{1}$ only weakly depends on $T$ except close to $T_N$. Meanwhile, as long as the concentration of Mn3 remains low, Mn3 can be treated as free spins in an external field (the molecular field induced by Mn1) with $M_3\propto \chi_{Mn3}(T) \propto 1/T$. Thus, upon cooling, the effective FM interaction increases much faster than the AF, resulting in a Ferri$^{AFM}$ to Ferri$^{FM}$ transition at $T_C<T_N$ once $|c_3J_{3}^{\rm{eff}}M_{3}|>c_1J_{1}M_{1}$. Furthermore, Figure 6(b) and the comparison of the neutron refinements of the $x=0$ \cite{ding2020crystal} and $x=0.76$ samples all indicate that $c_3$ increases and $c_1$ decreases with Sb doping. Therefore, inevitably, in the $T-x$ phase diagram, three doping regimes can appear: low doping region where the Ferri$^{AFM}$ state is
stable at any temperature, high doping region when only the Ferri$^{FM}$ state is stable,
and intermediate doping region, where the Ferri$^{FM}$ state is stable only up to some
$T_{C}<T_{N}$ with $T_C$ increasing with $x$. This doping dependence is exactly what we have observed in Fig. 1(k). 

We note that $l$ decreases with increasing $x$, so besides the formation of antisites, Sb doping also leads to the the magnetic dilution effect of the Mn1 sublattice. Future doping studies, such as Pb, Sn or Ge substitution on Mn site, if they will not cause antisite defects as the Sb doping does, they may provide a cleaner platform to investigate the magnetic dilution effect on the magnetism in the MBT family.

\begin{figure}
\centering
\includegraphics[width=3.4in]{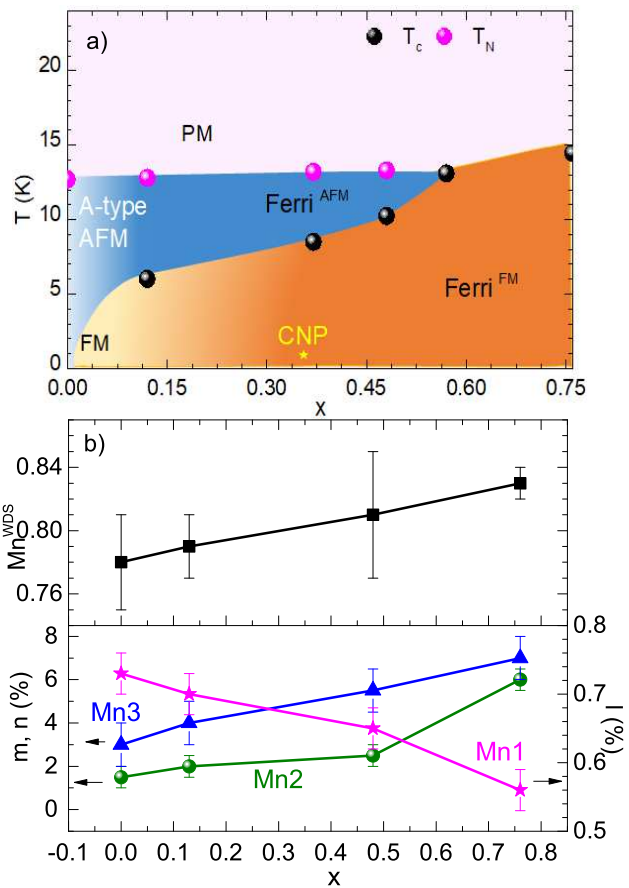} \caption{(a)  The temperature-doping ($T$ - $x$) phase diagram. With
increasing $x$, the carrier-type changes from electron to hole. A linear fitting with our data (Fig. S2) yields a charge neutrality
point (CNP) near $x$= 0.36. (b) The doping-dependent Mn$^{\rm{WDS}}$, $m$, $n$ and $l$.}
\label{fig:pd}
\end{figure}

Now let us turn our discussion to the band topology. It remains an open but
important question how robust the nontrivial band topology and thus the
associated emergent phenomena will be against the antisite defects. Recent
studies showed that Mn antisites are universal in MBT, $\sim$ 3(1)\% for
MnBi$_{2}$Te$_{4}$ and $\sim$ 13\%-16\% for MnSb$_{2}$Te$_{4}$
\cite{liu2021site,lai2021defect}. While it is impossible to construct a
structure model to reflect the real and complex chemical defects in our DFT calculations,
our attempts with the defect-free and periodic-Mn2-antisites scenarios shed light on
this puzzle. Two insightful observations can be made from Fig. 5. Firstly, in case II, the effect of Mn2 antisites is exaggerated
on the band topology, especially for the low and intermediate dopings where
the antisite defects are significantly smaller than 16.7\%. However, even
in this high-concentration case with periodic antisites, our calculations show robust band inversion and non-trivial topology, resulting in a
Weyl semimetal state at the intermediate dopings. Secondly, in the Bi-rich side, the characteristics of the band inversion are very similar for the two cases (Fig. 5(a) and (h)). However, in the Sb-rich side, the features of band inversion are apparently different (Fig. 5(e) and (k)). This observation implies that the antisite defect has stronger effect on modifying the band topology in the Sb-rich 147 phase than the Bi-rich 147 phase. 
Therefore, we believe that the non-trivial topology is likely robust
against the small amount of antisite defects here, especially at the Bi-rich side. Our finding is consistent
with the observation that despite 3(1)\% of antisite defects, a zero-field QAH effect appears at 1.5 K in a 5-SL device of MnBi$_{2}$Te$_{4}$
\cite{deng2020quantum}. Future
systematic ARPES measurements, in
combination with the DFT calculations on Sb doped MnBi$_4$Te$_7$,
will help settle this outstanding question.

\section{Conclusion} In Sb doped MnBi$_4$Te$_7$, the competition of the Mn1-Mn1 AFM interlayer coupling and the Mn3-assisted Mn1-Mn1 FM interlayer interaction leads to lower-temperature Ferri$^{FM}$ states and higher-temperature Ferri$^{AFM}$ states where
Mn3 sublattice is dynamically fluctuating at all temperatures. Meanwhile, the non-trivial band
topology appears robust against low or intermediate antisites, and points to a kaleidoscope of magnetic topological phases including FM axion
insulator state at low Sb-dopings and possible type-I ferrimagnetic Weyl
semimetal states at intermediate Sb dopings.

\section*{Acknowledgments}
NN thanks the useful discussions with A. P. Ramirez. We thank Randy Dumas at Quantum Design for measuring the isothermal magnetization up to 13 T. C. H. thanks the support
by the Julian Schwinger Fellowship at UCLA.
Work at UCLA was supported by the U.S. Department of Energy (DOE), Office of
Science, Office of Basic Energy Sciences under Award Number DE-SC0021117. T.-R.C. was supported by the Young Scholar Fellowship Program from the Ministry of Science and Technology (MOST) in Taiwan, under a MOST grant for the Columbus Program MOST110-2636-M-006-016, NCKU, Taiwan, and National Center for Theoretical Sciences, Taiwan. Work at NCKU was supported by the MOST, Taiwan, under grant MOST107-2627-E-006-001 and Higher Education Sprout Project, Ministry of Education to the Headquarters of University Advancement at NCKU. IM
acknowledges support from DOE under the grant DE-SC0021089. The work at ORNL was supported by the U.S. DOE, Office of Science, Office of Basic Energy Sciences, Early Career Research Program Award KC0402010, under Contract DE-AC05-00OR22725. This research used resources at the High Flux Isotope Reactor, the DOE Office of Science User Facility operated by ORNL.

\medskip

\bibliographystyle{apsrev4-1}
\bibliography{SbMBT}

\begin{thebibliography}{64}%
\makeatletter
\providecommand \@ifxundefined [1]{%
 \@ifx{#1\undefined}
}%
\providecommand \@ifnum [1]{%
 \ifnum #1\expandafter \@firstoftwo
 \else \expandafter \@secondoftwo
 \fi
}%
\providecommand \@ifx [1]{%
 \ifx #1\expandafter \@firstoftwo
 \else \expandafter \@secondoftwo
 \fi
}%
\providecommand \natexlab [1]{#1}%
\providecommand \enquote  [1]{``#1''}%
\providecommand \bibnamefont  [1]{#1}%
\providecommand \bibfnamefont [1]{#1}%
\providecommand \citenamefont [1]{#1}%
\providecommand \href@noop [0]{\@secondoftwo}%
\providecommand \href [0]{\begingroup \@sanitize@url \@href}%
\providecommand \@href[1]{\@@startlink{#1}\@@href}%
\providecommand \@@href[1]{\endgroup#1\@@endlink}%
\providecommand \@sanitize@url [0]{\catcode `\\12\catcode `\$12\catcode
  `\&12\catcode `\#12\catcode `\^12\catcode `\_12\catcode `\%12\relax}%
\providecommand \@@startlink[1]{}%
\providecommand \@@endlink[0]{}%
\providecommand \url  [0]{\begingroup\@sanitize@url \@url }%
\providecommand \@url [1]{\endgroup\@href {#1}{\urlprefix }}%
\providecommand \urlprefix  [0]{URL }%
\providecommand \Eprint [0]{\href }%
\providecommand \doibase [0]{http://dx.doi.org/}%
\providecommand \selectlanguage [0]{\@gobble}%
\providecommand \bibinfo  [0]{\@secondoftwo}%
\providecommand \bibfield  [0]{\@secondoftwo}%
\providecommand \translation [1]{[#1]}%
\providecommand \BibitemOpen [0]{}%
\providecommand \bibitemStop [0]{}%
\providecommand \bibitemNoStop [0]{.\EOS\space}%
\providecommand \EOS [0]{\spacefactor3000\relax}%
\providecommand \BibitemShut  [1]{\csname bibitem#1\endcsname}%
\let\auto@bib@innerbib\@empty
\bibitem [{\citenamefont {Tokura}\ \emph {et~al.}(2019)\citenamefont {Tokura},
  \citenamefont {Yasuda},\ and\ \citenamefont
  {Tsukazaki}}]{tokura2019magnetic}%
  \BibitemOpen
  \bibfield  {author} {\bibinfo {author} {\bibfnamefont {Y.}~\bibnamefont
  {Tokura}}, \bibinfo {author} {\bibfnamefont {K.}~\bibnamefont {Yasuda}}, \
  and\ \bibinfo {author} {\bibfnamefont {A.}~\bibnamefont {Tsukazaki}},\
  }\href@noop {} {\bibfield  {journal} {\bibinfo  {journal} {Nature Reviews
  Physics}\ }\textbf {\bibinfo {volume} {1}},\ \bibinfo {pages} {126} (\bibinfo
  {year} {2019})}\BibitemShut {NoStop}%
\bibitem [{\citenamefont {He}\ \emph {et~al.}(2018)\citenamefont {He},
  \citenamefont {Wang},\ and\ \citenamefont {Xue}}]{he2018topological}%
  \BibitemOpen
  \bibfield  {author} {\bibinfo {author} {\bibfnamefont {K.}~\bibnamefont
  {He}}, \bibinfo {author} {\bibfnamefont {Y.}~\bibnamefont {Wang}}, \ and\
  \bibinfo {author} {\bibfnamefont {Q.-K.}\ \bibnamefont {Xue}},\ }\href@noop
  {} {\bibfield  {journal} {\bibinfo  {journal} {Annual Review of Condensed
  Matter Physics}\ }\textbf {\bibinfo {volume} {9}},\ \bibinfo {pages} {329}
  (\bibinfo {year} {2018})}\BibitemShut {NoStop}%
\bibitem [{\citenamefont {Liu}\ \emph {et~al.}(2016)\citenamefont {Liu},
  \citenamefont {Zhang},\ and\ \citenamefont {Qi}}]{liu2016quantum}%
  \BibitemOpen
  \bibfield  {author} {\bibinfo {author} {\bibfnamefont {C.-X.}\ \bibnamefont
  {Liu}}, \bibinfo {author} {\bibfnamefont {S.-C.}\ \bibnamefont {Zhang}}, \
  and\ \bibinfo {author} {\bibfnamefont {X.-L.}\ \bibnamefont {Qi}},\
  }\href@noop {} {\bibfield  {journal} {\bibinfo  {journal} {Annual Review of
  Condensed Matter Physics}\ }\textbf {\bibinfo {volume} {7}},\ \bibinfo
  {pages} {301} (\bibinfo {year} {2016})}\BibitemShut {NoStop}%
\bibitem [{\citenamefont {Wang}\ \emph {et~al.}(2015)\citenamefont {Wang},
  \citenamefont {Lian}, \citenamefont {Qi},\ and\ \citenamefont
  {Zhang}}]{wang2015quantized}%
  \BibitemOpen
  \bibfield  {author} {\bibinfo {author} {\bibfnamefont {J.}~\bibnamefont
  {Wang}}, \bibinfo {author} {\bibfnamefont {B.}~\bibnamefont {Lian}}, \bibinfo
  {author} {\bibfnamefont {X.-L.}\ \bibnamefont {Qi}}, \ and\ \bibinfo {author}
  {\bibfnamefont {S.-C.}\ \bibnamefont {Zhang}},\ }\href@noop {} {\bibfield
  {journal} {\bibinfo  {journal} {Physical Review B}\ }\textbf {\bibinfo
  {volume} {92}},\ \bibinfo {pages} {081107} (\bibinfo {year}
  {2015})}\BibitemShut {NoStop}%
\bibitem [{\citenamefont {Lee}\ \emph {et~al.}(2013)\citenamefont {Lee},
  \citenamefont {Kim}, \citenamefont {Park}, \citenamefont {Chung},
  \citenamefont {Lim}, \citenamefont {Seo},\ and\ \citenamefont
  {Park}}]{lee2013crystal}%
  \BibitemOpen
  \bibfield  {author} {\bibinfo {author} {\bibfnamefont {D.~S.}\ \bibnamefont
  {Lee}}, \bibinfo {author} {\bibfnamefont {T.-H.}\ \bibnamefont {Kim}},
  \bibinfo {author} {\bibfnamefont {C.-H.}\ \bibnamefont {Park}}, \bibinfo
  {author} {\bibfnamefont {C.-Y.}\ \bibnamefont {Chung}}, \bibinfo {author}
  {\bibfnamefont {Y.~S.}\ \bibnamefont {Lim}}, \bibinfo {author} {\bibfnamefont
  {W.-S.}\ \bibnamefont {Seo}}, \ and\ \bibinfo {author} {\bibfnamefont
  {H.-H.}\ \bibnamefont {Park}},\ }\href@noop {} {\bibfield  {journal}
  {\bibinfo  {journal} {CrystEngComm}\ }\textbf {\bibinfo {volume} {15}},\
  \bibinfo {pages} {5532} (\bibinfo {year} {2013})}\BibitemShut {NoStop}%
\bibitem [{\citenamefont {Rienks}\ \emph {et~al.}(2019)\citenamefont {Rienks},
  \citenamefont {Wimmer}, \citenamefont {S{\'a}nchez~Barriga}, \citenamefont
  {Caha}, \citenamefont {Mandal}, \citenamefont {R{\r{u}}{\v{z}}i{\v{c}}ka},
  \citenamefont {Ney}, \citenamefont {Steiner}, \citenamefont {Volobuev},
  \citenamefont {Groiss} \emph {et~al.}}]{rienks2019large}%
  \BibitemOpen
  \bibfield  {author} {\bibinfo {author} {\bibfnamefont {E.}~\bibnamefont
  {Rienks}}, \bibinfo {author} {\bibfnamefont {S.}~\bibnamefont {Wimmer}},
  \bibinfo {author} {\bibfnamefont {J.}~\bibnamefont {S{\'a}nchez~Barriga}},
  \bibinfo {author} {\bibfnamefont {O.}~\bibnamefont {Caha}}, \bibinfo {author}
  {\bibfnamefont {P.}~\bibnamefont {Mandal}}, \bibinfo {author} {\bibfnamefont
  {J.}~\bibnamefont {R{\r{u}}{\v{z}}i{\v{c}}ka}}, \bibinfo {author}
  {\bibfnamefont {A.}~\bibnamefont {Ney}}, \bibinfo {author} {\bibfnamefont
  {H.}~\bibnamefont {Steiner}}, \bibinfo {author} {\bibfnamefont
  {V.}~\bibnamefont {Volobuev}}, \bibinfo {author} {\bibfnamefont
  {H.}~\bibnamefont {Groiss}},  \emph {et~al.},\ }\href@noop {} {\bibfield
  {journal} {\bibinfo  {journal} {Nature}\ }\textbf {\bibinfo {volume} {576}},\
  \bibinfo {pages} {423} (\bibinfo {year} {2019})}\BibitemShut {NoStop}%
\bibitem [{\citenamefont {Zhang}\ \emph {et~al.}(2019)\citenamefont {Zhang},
  \citenamefont {Shi}, \citenamefont {Zhu}, \citenamefont {Xing}, \citenamefont
  {Zhang},\ and\ \citenamefont {Wang}}]{zhang2019topological}%
  \BibitemOpen
  \bibfield  {author} {\bibinfo {author} {\bibfnamefont {D.}~\bibnamefont
  {Zhang}}, \bibinfo {author} {\bibfnamefont {M.}~\bibnamefont {Shi}}, \bibinfo
  {author} {\bibfnamefont {T.}~\bibnamefont {Zhu}}, \bibinfo {author}
  {\bibfnamefont {D.}~\bibnamefont {Xing}}, \bibinfo {author} {\bibfnamefont
  {H.}~\bibnamefont {Zhang}}, \ and\ \bibinfo {author} {\bibfnamefont
  {J.}~\bibnamefont {Wang}},\ }\href@noop {} {\bibfield  {journal} {\bibinfo
  {journal} {Physical review letters}\ }\textbf {\bibinfo {volume} {122}},\
  \bibinfo {pages} {206401} (\bibinfo {year} {2019})}\BibitemShut {NoStop}%
\bibitem [{\citenamefont {Li}\ \emph {et~al.}(2019{\natexlab{a}})\citenamefont
  {Li}, \citenamefont {Li}, \citenamefont {Du}, \citenamefont {Wang},
  \citenamefont {Gu}, \citenamefont {Zhang}, \citenamefont {He}, \citenamefont
  {Duan},\ and\ \citenamefont {Xu}}]{li2019intrinsic}%
  \BibitemOpen
  \bibfield  {author} {\bibinfo {author} {\bibfnamefont {J.}~\bibnamefont
  {Li}}, \bibinfo {author} {\bibfnamefont {Y.}~\bibnamefont {Li}}, \bibinfo
  {author} {\bibfnamefont {S.}~\bibnamefont {Du}}, \bibinfo {author}
  {\bibfnamefont {Z.}~\bibnamefont {Wang}}, \bibinfo {author} {\bibfnamefont
  {B.-L.}\ \bibnamefont {Gu}}, \bibinfo {author} {\bibfnamefont {S.-C.}\
  \bibnamefont {Zhang}}, \bibinfo {author} {\bibfnamefont {K.}~\bibnamefont
  {He}}, \bibinfo {author} {\bibfnamefont {W.}~\bibnamefont {Duan}}, \ and\
  \bibinfo {author} {\bibfnamefont {Y.}~\bibnamefont {Xu}},\ }\href@noop {}
  {\bibfield  {journal} {\bibinfo  {journal} {Science Advances}\ }\textbf
  {\bibinfo {volume} {5}},\ \bibinfo {pages} {eaaw5685} (\bibinfo {year}
  {2019}{\natexlab{a}})}\BibitemShut {NoStop}%
\bibitem [{\citenamefont {Otrokov}\ \emph
  {et~al.}(2019{\natexlab{a}})\citenamefont {Otrokov}, \citenamefont
  {Klimovskikh}, \citenamefont {Bentmann}, \citenamefont {Estyunin},
  \citenamefont {Zeugner}, \citenamefont {Aliev}, \citenamefont {Ga{\ss}},
  \citenamefont {Wolter}, \citenamefont {Koroleva}, \citenamefont {Shikin}
  \emph {et~al.}}]{otrokov2019prediction}%
  \BibitemOpen
  \bibfield  {author} {\bibinfo {author} {\bibfnamefont {M.~M.}\ \bibnamefont
  {Otrokov}}, \bibinfo {author} {\bibfnamefont {I.~I.}\ \bibnamefont
  {Klimovskikh}}, \bibinfo {author} {\bibfnamefont {H.}~\bibnamefont
  {Bentmann}}, \bibinfo {author} {\bibfnamefont {D.}~\bibnamefont {Estyunin}},
  \bibinfo {author} {\bibfnamefont {A.}~\bibnamefont {Zeugner}}, \bibinfo
  {author} {\bibfnamefont {Z.~S.}\ \bibnamefont {Aliev}}, \bibinfo {author}
  {\bibfnamefont {S.}~\bibnamefont {Ga{\ss}}}, \bibinfo {author} {\bibfnamefont
  {A.}~\bibnamefont {Wolter}}, \bibinfo {author} {\bibfnamefont
  {A.}~\bibnamefont {Koroleva}}, \bibinfo {author} {\bibfnamefont {A.~M.}\
  \bibnamefont {Shikin}},  \emph {et~al.},\ }\href@noop {} {\bibfield
  {journal} {\bibinfo  {journal} {Nature}\ }\textbf {\bibinfo {volume} {576}},\
  \bibinfo {pages} {416} (\bibinfo {year} {2019}{\natexlab{a}})}\BibitemShut
  {NoStop}%
\bibitem [{\citenamefont {Aliev}\ \emph {et~al.}(2019)\citenamefont {Aliev},
  \citenamefont {Amiraslanov}, \citenamefont {Nasonova}, \citenamefont
  {Shevelkov}, \citenamefont {Abdullayev}, \citenamefont {Jahangirli},
  \citenamefont {Orujlu}, \citenamefont {Otrokov}, \citenamefont {Mamedov},
  \citenamefont {Babanly} \emph {et~al.}}]{aliev2019novel}%
  \BibitemOpen
  \bibfield  {author} {\bibinfo {author} {\bibfnamefont {Z.~S.}\ \bibnamefont
  {Aliev}}, \bibinfo {author} {\bibfnamefont {I.~R.}\ \bibnamefont
  {Amiraslanov}}, \bibinfo {author} {\bibfnamefont {D.~I.}\ \bibnamefont
  {Nasonova}}, \bibinfo {author} {\bibfnamefont {A.~V.}\ \bibnamefont
  {Shevelkov}}, \bibinfo {author} {\bibfnamefont {N.~A.}\ \bibnamefont
  {Abdullayev}}, \bibinfo {author} {\bibfnamefont {Z.~A.}\ \bibnamefont
  {Jahangirli}}, \bibinfo {author} {\bibfnamefont {E.~N.}\ \bibnamefont
  {Orujlu}}, \bibinfo {author} {\bibfnamefont {M.~M.}\ \bibnamefont {Otrokov}},
  \bibinfo {author} {\bibfnamefont {N.~T.}\ \bibnamefont {Mamedov}}, \bibinfo
  {author} {\bibfnamefont {M.~B.}\ \bibnamefont {Babanly}},  \emph {et~al.},\
  }\href@noop {} {\bibfield  {journal} {\bibinfo  {journal} {Journal of Alloys
  and Compounds}\ }\textbf {\bibinfo {volume} {789}},\ \bibinfo {pages} {443}
  (\bibinfo {year} {2019})}\BibitemShut {NoStop}%
\bibitem [{\citenamefont {Hu}\ \emph {et~al.}(2020{\natexlab{a}})\citenamefont
  {Hu}, \citenamefont {Gordon}, \citenamefont {Liu}, \citenamefont {Liu},
  \citenamefont {Zhou}, \citenamefont {Hao}, \citenamefont {Narayan},
  \citenamefont {Emmanouilidou}, \citenamefont {Sun}, \citenamefont {Liu} \emph
  {et~al.}}]{147}%
  \BibitemOpen
  \bibfield  {author} {\bibinfo {author} {\bibfnamefont {C.}~\bibnamefont
  {Hu}}, \bibinfo {author} {\bibfnamefont {K.~N.}\ \bibnamefont {Gordon}},
  \bibinfo {author} {\bibfnamefont {P.}~\bibnamefont {Liu}}, \bibinfo {author}
  {\bibfnamefont {J.}~\bibnamefont {Liu}}, \bibinfo {author} {\bibfnamefont
  {X.}~\bibnamefont {Zhou}}, \bibinfo {author} {\bibfnamefont {P.}~\bibnamefont
  {Hao}}, \bibinfo {author} {\bibfnamefont {D.}~\bibnamefont {Narayan}},
  \bibinfo {author} {\bibfnamefont {E.}~\bibnamefont {Emmanouilidou}}, \bibinfo
  {author} {\bibfnamefont {H.}~\bibnamefont {Sun}}, \bibinfo {author}
  {\bibfnamefont {Y.}~\bibnamefont {Liu}},  \emph {et~al.},\ }\href@noop {}
  {\bibfield  {journal} {\bibinfo  {journal} {Nature communications}\ }\textbf
  {\bibinfo {volume} {11}},\ \bibinfo {pages} {1} (\bibinfo {year}
  {2020}{\natexlab{a}})}\BibitemShut {NoStop}%
\bibitem [{\citenamefont {Hu}\ \emph {et~al.}(2020{\natexlab{b}})\citenamefont
  {Hu}, \citenamefont {Ding}, \citenamefont {Gordon}, \citenamefont {Ghosh},
  \citenamefont {Tien}, \citenamefont {Li}, \citenamefont {Linn}, \citenamefont
  {Lien}, \citenamefont {Huang}, \citenamefont {Mackey} \emph {et~al.}}]{1813}%
  \BibitemOpen
  \bibfield  {author} {\bibinfo {author} {\bibfnamefont {C.}~\bibnamefont
  {Hu}}, \bibinfo {author} {\bibfnamefont {L.}~\bibnamefont {Ding}}, \bibinfo
  {author} {\bibfnamefont {K.~N.}\ \bibnamefont {Gordon}}, \bibinfo {author}
  {\bibfnamefont {B.}~\bibnamefont {Ghosh}}, \bibinfo {author} {\bibfnamefont
  {H.-J.}\ \bibnamefont {Tien}}, \bibinfo {author} {\bibfnamefont
  {H.}~\bibnamefont {Li}}, \bibinfo {author} {\bibfnamefont {A.~G.}\
  \bibnamefont {Linn}}, \bibinfo {author} {\bibfnamefont {S.-W.}\ \bibnamefont
  {Lien}}, \bibinfo {author} {\bibfnamefont {C.-Y.}\ \bibnamefont {Huang}},
  \bibinfo {author} {\bibfnamefont {S.}~\bibnamefont {Mackey}},  \emph
  {et~al.},\ }\href@noop {} {\bibfield  {journal} {\bibinfo  {journal} {Science
  Advances}\ }\textbf {\bibinfo {volume} {6}},\ \bibinfo {pages} {eaba4275}
  (\bibinfo {year} {2020}{\natexlab{b}})}\BibitemShut {NoStop}%
\bibitem [{\citenamefont {Deng}\ \emph
  {et~al.}(2020{\natexlab{a}})\citenamefont {Deng}, \citenamefont {Chen},
  \citenamefont {Wo{\l{l}}o{\'{s}}}, \citenamefont {Konczykowski},
  \citenamefont {Sobczak}, \citenamefont {Sitnicka}, \citenamefont
  {Fedorchenko}, \citenamefont {Borysiuk}, \citenamefont {Heider},
  \citenamefont {Pluci{\'{n}}ski}, \citenamefont {Park}, \citenamefont
  {Georgescu}, \citenamefont {Cano},\ and\ \citenamefont
  {Krusin~Elbaum}}]{deng2020high-temperature}%
  \BibitemOpen
  \bibfield  {author} {\bibinfo {author} {\bibfnamefont {H.}~\bibnamefont
  {Deng}}, \bibinfo {author} {\bibfnamefont {Z.}~\bibnamefont {Chen}}, \bibinfo
  {author} {\bibfnamefont {A.}~\bibnamefont {Wo{\l{l}}o{\'{s}}}}, \bibinfo
  {author} {\bibfnamefont {M.}~\bibnamefont {Konczykowski}}, \bibinfo {author}
  {\bibfnamefont {K.}~\bibnamefont {Sobczak}}, \bibinfo {author} {\bibfnamefont
  {J.}~\bibnamefont {Sitnicka}}, \bibinfo {author} {\bibfnamefont {I.~V.}\
  \bibnamefont {Fedorchenko}}, \bibinfo {author} {\bibfnamefont
  {J.}~\bibnamefont {Borysiuk}}, \bibinfo {author} {\bibfnamefont
  {T.}~\bibnamefont {Heider}}, \bibinfo {author} {\bibfnamefont
  {{\l{L}}.}~\bibnamefont {Pluci{\'{n}}ski}}, \bibinfo {author} {\bibfnamefont
  {K.}~\bibnamefont {Park}}, \bibinfo {author} {\bibfnamefont {A.~B.}\
  \bibnamefont {Georgescu}}, \bibinfo {author} {\bibfnamefont {J.}~\bibnamefont
  {Cano}}, \ and\ \bibinfo {author} {\bibfnamefont {L.}~\bibnamefont
  {Krusin~Elbaum}},\ }\href {\doibase 10.1038/s41567-020-0998-2} {\bibfield
  {journal} {\bibinfo  {journal} {Nature Physics}\ } (\bibinfo {year}
  {2020}{\natexlab{a}}),\ 10.1038/s41567-020-0998-2}\BibitemShut {NoStop}%
\bibitem [{\citenamefont {Gong}\ \emph {et~al.}(2019)\citenamefont {Gong},
  \citenamefont {Guo}, \citenamefont {Li}, \citenamefont {Zhu}, \citenamefont
  {Liao}, \citenamefont {Liu}, \citenamefont {Zhang}, \citenamefont {Gu},
  \citenamefont {Tang}, \citenamefont {Feng} \emph
  {et~al.}}]{gong2019experimental}%
  \BibitemOpen
  \bibfield  {author} {\bibinfo {author} {\bibfnamefont {Y.}~\bibnamefont
  {Gong}}, \bibinfo {author} {\bibfnamefont {J.}~\bibnamefont {Guo}}, \bibinfo
  {author} {\bibfnamefont {J.}~\bibnamefont {Li}}, \bibinfo {author}
  {\bibfnamefont {K.}~\bibnamefont {Zhu}}, \bibinfo {author} {\bibfnamefont
  {M.}~\bibnamefont {Liao}}, \bibinfo {author} {\bibfnamefont {X.}~\bibnamefont
  {Liu}}, \bibinfo {author} {\bibfnamefont {Q.}~\bibnamefont {Zhang}}, \bibinfo
  {author} {\bibfnamefont {L.}~\bibnamefont {Gu}}, \bibinfo {author}
  {\bibfnamefont {L.}~\bibnamefont {Tang}}, \bibinfo {author} {\bibfnamefont
  {X.}~\bibnamefont {Feng}},  \emph {et~al.},\ }\href@noop {} {\bibfield
  {journal} {\bibinfo  {journal} {Chinese Physics Letters}\ }\textbf {\bibinfo
  {volume} {36}},\ \bibinfo {pages} {076801} (\bibinfo {year}
  {2019})}\BibitemShut {NoStop}%
\bibitem [{\citenamefont {Lee}\ \emph {et~al.}(2019)\citenamefont {Lee},
  \citenamefont {Zhu}, \citenamefont {Wang}, \citenamefont {Miao},
  \citenamefont {Pillsbury}, \citenamefont {Yi}, \citenamefont {Kempinger},
  \citenamefont {Hu}, \citenamefont {Heikes}, \citenamefont {Quarterman} \emph
  {et~al.}}]{lee2019spin}%
  \BibitemOpen
  \bibfield  {author} {\bibinfo {author} {\bibfnamefont {S.~H.}\ \bibnamefont
  {Lee}}, \bibinfo {author} {\bibfnamefont {Y.}~\bibnamefont {Zhu}}, \bibinfo
  {author} {\bibfnamefont {Y.}~\bibnamefont {Wang}}, \bibinfo {author}
  {\bibfnamefont {L.}~\bibnamefont {Miao}}, \bibinfo {author} {\bibfnamefont
  {T.}~\bibnamefont {Pillsbury}}, \bibinfo {author} {\bibfnamefont
  {H.}~\bibnamefont {Yi}}, \bibinfo {author} {\bibfnamefont {S.}~\bibnamefont
  {Kempinger}}, \bibinfo {author} {\bibfnamefont {J.}~\bibnamefont {Hu}},
  \bibinfo {author} {\bibfnamefont {C.~A.}\ \bibnamefont {Heikes}}, \bibinfo
  {author} {\bibfnamefont {P.}~\bibnamefont {Quarterman}},  \emph {et~al.},\
  }\href@noop {} {\bibfield  {journal} {\bibinfo  {journal} {Physical Review
  Research}\ }\textbf {\bibinfo {volume} {1}},\ \bibinfo {pages} {012011}
  (\bibinfo {year} {2019})}\BibitemShut {NoStop}%
\bibitem [{\citenamefont {Yan}\ \emph {et~al.}(2019{\natexlab{a}})\citenamefont
  {Yan}, \citenamefont {Zhang}, \citenamefont {Heitmann}, \citenamefont
  {Huang}, \citenamefont {Chen}, \citenamefont {Cheng}, \citenamefont {Wu},
  \citenamefont {Vaknin}, \citenamefont {Sales},\ and\ \citenamefont
  {McQueeney}}]{yan2019crystal}%
  \BibitemOpen
  \bibfield  {author} {\bibinfo {author} {\bibfnamefont {J.-Q.}\ \bibnamefont
  {Yan}}, \bibinfo {author} {\bibfnamefont {Q.}~\bibnamefont {Zhang}}, \bibinfo
  {author} {\bibfnamefont {T.}~\bibnamefont {Heitmann}}, \bibinfo {author}
  {\bibfnamefont {Z.}~\bibnamefont {Huang}}, \bibinfo {author} {\bibfnamefont
  {K.}~\bibnamefont {Chen}}, \bibinfo {author} {\bibfnamefont {J.-G.}\
  \bibnamefont {Cheng}}, \bibinfo {author} {\bibfnamefont {W.}~\bibnamefont
  {Wu}}, \bibinfo {author} {\bibfnamefont {D.}~\bibnamefont {Vaknin}}, \bibinfo
  {author} {\bibfnamefont {B.~C.}\ \bibnamefont {Sales}}, \ and\ \bibinfo
  {author} {\bibfnamefont {R.~J.}\ \bibnamefont {McQueeney}},\ }\href@noop {}
  {\bibfield  {journal} {\bibinfo  {journal} {Physical Review Materials}\
  }\textbf {\bibinfo {volume} {3}},\ \bibinfo {pages} {064202} (\bibinfo {year}
  {2019}{\natexlab{a}})}\BibitemShut {NoStop}%
\bibitem [{\citenamefont {Zeugner}\ \emph {et~al.}(2019)\citenamefont
  {Zeugner}, \citenamefont {Nietschke}, \citenamefont {Wolter}, \citenamefont
  {Ga{\ss}}, \citenamefont {Vidal}, \citenamefont {Peixoto}, \citenamefont
  {Pohl}, \citenamefont {Damm}, \citenamefont {Lubk}, \citenamefont {Hentrich}
  \emph {et~al.}}]{zeugner2019chemical}%
  \BibitemOpen
  \bibfield  {author} {\bibinfo {author} {\bibfnamefont {A.}~\bibnamefont
  {Zeugner}}, \bibinfo {author} {\bibfnamefont {F.}~\bibnamefont {Nietschke}},
  \bibinfo {author} {\bibfnamefont {A.~U.}\ \bibnamefont {Wolter}}, \bibinfo
  {author} {\bibfnamefont {S.}~\bibnamefont {Ga{\ss}}}, \bibinfo {author}
  {\bibfnamefont {R.~C.}\ \bibnamefont {Vidal}}, \bibinfo {author}
  {\bibfnamefont {T.~R.}\ \bibnamefont {Peixoto}}, \bibinfo {author}
  {\bibfnamefont {D.}~\bibnamefont {Pohl}}, \bibinfo {author} {\bibfnamefont
  {C.}~\bibnamefont {Damm}}, \bibinfo {author} {\bibfnamefont {A.}~\bibnamefont
  {Lubk}}, \bibinfo {author} {\bibfnamefont {R.}~\bibnamefont {Hentrich}},
  \emph {et~al.},\ }\href@noop {} {\bibfield  {journal} {\bibinfo  {journal}
  {Chemistry of Materials}\ }\textbf {\bibinfo {volume} {31}},\ \bibinfo
  {pages} {2795} (\bibinfo {year} {2019})}\BibitemShut {NoStop}%
\bibitem [{\citenamefont {Otrokov}\ \emph
  {et~al.}(2019{\natexlab{b}})\citenamefont {Otrokov}, \citenamefont {Rusinov},
  \citenamefont {Blanco-Rey}, \citenamefont {Hoffmann}, \citenamefont
  {Vyazovskaya}, \citenamefont {Eremeev}, \citenamefont {Ernst}, \citenamefont
  {Echenique}, \citenamefont {Arnau},\ and\ \citenamefont
  {Chulkov}}]{otrokov2019unique}%
  \BibitemOpen
  \bibfield  {author} {\bibinfo {author} {\bibfnamefont {M.~M.}\ \bibnamefont
  {Otrokov}}, \bibinfo {author} {\bibfnamefont {I.~P.}\ \bibnamefont
  {Rusinov}}, \bibinfo {author} {\bibfnamefont {M.}~\bibnamefont {Blanco-Rey}},
  \bibinfo {author} {\bibfnamefont {M.}~\bibnamefont {Hoffmann}}, \bibinfo
  {author} {\bibfnamefont {A.~Y.}\ \bibnamefont {Vyazovskaya}}, \bibinfo
  {author} {\bibfnamefont {S.~V.}\ \bibnamefont {Eremeev}}, \bibinfo {author}
  {\bibfnamefont {A.}~\bibnamefont {Ernst}}, \bibinfo {author} {\bibfnamefont
  {P.~M.}\ \bibnamefont {Echenique}}, \bibinfo {author} {\bibfnamefont
  {A.}~\bibnamefont {Arnau}}, \ and\ \bibinfo {author} {\bibfnamefont {E.~V.}\
  \bibnamefont {Chulkov}},\ }\href@noop {} {\bibfield  {journal} {\bibinfo
  {journal} {Physical Review Letters}\ }\textbf {\bibinfo {volume} {122}},\
  \bibinfo {pages} {107202} (\bibinfo {year} {2019}{\natexlab{b}})}\BibitemShut
  {NoStop}%
\bibitem [{\citenamefont {Chen}\ \emph
  {et~al.}(2019{\natexlab{a}})\citenamefont {Chen}, \citenamefont {Fei},
  \citenamefont {Zhang}, \citenamefont {Zhang}, \citenamefont {Liu},
  \citenamefont {Zhang}, \citenamefont {Wang}, \citenamefont {Wei},
  \citenamefont {Zhang}, \citenamefont {Zuo} \emph
  {et~al.}}]{chen2019intrinsic}%
  \BibitemOpen
  \bibfield  {author} {\bibinfo {author} {\bibfnamefont {B.}~\bibnamefont
  {Chen}}, \bibinfo {author} {\bibfnamefont {F.}~\bibnamefont {Fei}}, \bibinfo
  {author} {\bibfnamefont {D.}~\bibnamefont {Zhang}}, \bibinfo {author}
  {\bibfnamefont {B.}~\bibnamefont {Zhang}}, \bibinfo {author} {\bibfnamefont
  {W.}~\bibnamefont {Liu}}, \bibinfo {author} {\bibfnamefont {S.}~\bibnamefont
  {Zhang}}, \bibinfo {author} {\bibfnamefont {P.}~\bibnamefont {Wang}},
  \bibinfo {author} {\bibfnamefont {B.}~\bibnamefont {Wei}}, \bibinfo {author}
  {\bibfnamefont {Y.}~\bibnamefont {Zhang}}, \bibinfo {author} {\bibfnamefont
  {Z.}~\bibnamefont {Zuo}},  \emph {et~al.},\ }\href@noop {} {\bibfield
  {journal} {\bibinfo  {journal} {Nature communications}\ }\textbf {\bibinfo
  {volume} {10}},\ \bibinfo {pages} {1} (\bibinfo {year}
  {2019}{\natexlab{a}})}\BibitemShut {NoStop}%
\bibitem [{\citenamefont {Wu}\ \emph {et~al.}(2019)\citenamefont {Wu},
  \citenamefont {Liu}, \citenamefont {Sasase}, \citenamefont {Ienaga},
  \citenamefont {Obata}, \citenamefont {Yukawa}, \citenamefont {Horiba},
  \citenamefont {Kumigashira}, \citenamefont {Okuma}, \citenamefont {Inoshita}
  \emph {et~al.}}]{wu2019natural}%
  \BibitemOpen
  \bibfield  {author} {\bibinfo {author} {\bibfnamefont {J.}~\bibnamefont
  {Wu}}, \bibinfo {author} {\bibfnamefont {F.}~\bibnamefont {Liu}}, \bibinfo
  {author} {\bibfnamefont {M.}~\bibnamefont {Sasase}}, \bibinfo {author}
  {\bibfnamefont {K.}~\bibnamefont {Ienaga}}, \bibinfo {author} {\bibfnamefont
  {Y.}~\bibnamefont {Obata}}, \bibinfo {author} {\bibfnamefont
  {R.}~\bibnamefont {Yukawa}}, \bibinfo {author} {\bibfnamefont
  {K.}~\bibnamefont {Horiba}}, \bibinfo {author} {\bibfnamefont
  {H.}~\bibnamefont {Kumigashira}}, \bibinfo {author} {\bibfnamefont
  {S.}~\bibnamefont {Okuma}}, \bibinfo {author} {\bibfnamefont
  {T.}~\bibnamefont {Inoshita}},  \emph {et~al.},\ }\href@noop {} {\bibfield
  {journal} {\bibinfo  {journal} {Science advances}\ }\textbf {\bibinfo
  {volume} {5}},\ \bibinfo {pages} {eaax9989} (\bibinfo {year}
  {2019})}\BibitemShut {NoStop}%
\bibitem [{\citenamefont {Hao}\ \emph {et~al.}(2019)\citenamefont {Hao},
  \citenamefont {Liu}, \citenamefont {Feng}, \citenamefont {Ma}, \citenamefont
  {Schwier}, \citenamefont {Arita}, \citenamefont {Kumar}, \citenamefont {Hu},
  \citenamefont {Zeng}, \citenamefont {Wang} \emph {et~al.}}]{hao2019gapless}%
  \BibitemOpen
  \bibfield  {author} {\bibinfo {author} {\bibfnamefont {Y.-J.}\ \bibnamefont
  {Hao}}, \bibinfo {author} {\bibfnamefont {P.}~\bibnamefont {Liu}}, \bibinfo
  {author} {\bibfnamefont {Y.}~\bibnamefont {Feng}}, \bibinfo {author}
  {\bibfnamefont {X.-M.}\ \bibnamefont {Ma}}, \bibinfo {author} {\bibfnamefont
  {E.~F.}\ \bibnamefont {Schwier}}, \bibinfo {author} {\bibfnamefont
  {M.}~\bibnamefont {Arita}}, \bibinfo {author} {\bibfnamefont
  {S.}~\bibnamefont {Kumar}}, \bibinfo {author} {\bibfnamefont
  {C.}~\bibnamefont {Hu}}, \bibinfo {author} {\bibfnamefont {M.}~\bibnamefont
  {Zeng}}, \bibinfo {author} {\bibfnamefont {Y.}~\bibnamefont {Wang}},  \emph
  {et~al.},\ }\href@noop {} {\bibfield  {journal} {\bibinfo  {journal}
  {Physical Review X}\ }\textbf {\bibinfo {volume} {9}},\ \bibinfo {pages}
  {041038} (\bibinfo {year} {2019})}\BibitemShut {NoStop}%
\bibitem [{\citenamefont {Deng}\ \emph
  {et~al.}(2020{\natexlab{b}})\citenamefont {Deng}, \citenamefont {Yu},
  \citenamefont {Shi}, \citenamefont {Guo}, \citenamefont {Xu}, \citenamefont
  {Wang}, \citenamefont {Chen},\ and\ \citenamefont {Zhang}}]{deng2020quantum}%
  \BibitemOpen
  \bibfield  {author} {\bibinfo {author} {\bibfnamefont {Y.}~\bibnamefont
  {Deng}}, \bibinfo {author} {\bibfnamefont {Y.}~\bibnamefont {Yu}}, \bibinfo
  {author} {\bibfnamefont {M.~Z.}\ \bibnamefont {Shi}}, \bibinfo {author}
  {\bibfnamefont {Z.}~\bibnamefont {Guo}}, \bibinfo {author} {\bibfnamefont
  {Z.}~\bibnamefont {Xu}}, \bibinfo {author} {\bibfnamefont {J.}~\bibnamefont
  {Wang}}, \bibinfo {author} {\bibfnamefont {X.~H.}\ \bibnamefont {Chen}}, \
  and\ \bibinfo {author} {\bibfnamefont {Y.}~\bibnamefont {Zhang}},\
  }\href@noop {} {\bibfield  {journal} {\bibinfo  {journal} {Science}\ }\textbf
  {\bibinfo {volume} {367}},\ \bibinfo {pages} {895} (\bibinfo {year}
  {2020}{\natexlab{b}})}\BibitemShut {NoStop}%
\bibitem [{\citenamefont {Liu}\ \emph {et~al.}(2020)\citenamefont {Liu},
  \citenamefont {Wang}, \citenamefont {Li}, \citenamefont {Wu}, \citenamefont
  {Li}, \citenamefont {Li}, \citenamefont {He}, \citenamefont {Xu},
  \citenamefont {Zhang},\ and\ \citenamefont {Wang}}]{liu2020robust}%
  \BibitemOpen
  \bibfield  {author} {\bibinfo {author} {\bibfnamefont {C.}~\bibnamefont
  {Liu}}, \bibinfo {author} {\bibfnamefont {Y.}~\bibnamefont {Wang}}, \bibinfo
  {author} {\bibfnamefont {H.}~\bibnamefont {Li}}, \bibinfo {author}
  {\bibfnamefont {Y.}~\bibnamefont {Wu}}, \bibinfo {author} {\bibfnamefont
  {Y.}~\bibnamefont {Li}}, \bibinfo {author} {\bibfnamefont {J.}~\bibnamefont
  {Li}}, \bibinfo {author} {\bibfnamefont {K.}~\bibnamefont {He}}, \bibinfo
  {author} {\bibfnamefont {Y.}~\bibnamefont {Xu}}, \bibinfo {author}
  {\bibfnamefont {J.}~\bibnamefont {Zhang}}, \ and\ \bibinfo {author}
  {\bibfnamefont {Y.}~\bibnamefont {Wang}},\ }\href@noop {} {\bibfield
  {journal} {\bibinfo  {journal} {Nature Materials}\ }\textbf {\bibinfo
  {volume} {19}},\ \bibinfo {pages} {522} (\bibinfo {year} {2020})}\BibitemShut
  {NoStop}%
\bibitem [{\citenamefont {Chen}\ \emph
  {et~al.}(2019{\natexlab{b}})\citenamefont {Chen}, \citenamefont {Xu},
  \citenamefont {Li}, \citenamefont {Li}, \citenamefont {Wang}, \citenamefont
  {Zhang}, \citenamefont {Li}, \citenamefont {Wu}, \citenamefont {Liang},
  \citenamefont {Chen} \emph {et~al.}}]{chen2019topological}%
  \BibitemOpen
  \bibfield  {author} {\bibinfo {author} {\bibfnamefont {Y.}~\bibnamefont
  {Chen}}, \bibinfo {author} {\bibfnamefont {L.}~\bibnamefont {Xu}}, \bibinfo
  {author} {\bibfnamefont {J.}~\bibnamefont {Li}}, \bibinfo {author}
  {\bibfnamefont {Y.}~\bibnamefont {Li}}, \bibinfo {author} {\bibfnamefont
  {H.}~\bibnamefont {Wang}}, \bibinfo {author} {\bibfnamefont {C.}~\bibnamefont
  {Zhang}}, \bibinfo {author} {\bibfnamefont {H.}~\bibnamefont {Li}}, \bibinfo
  {author} {\bibfnamefont {Y.}~\bibnamefont {Wu}}, \bibinfo {author}
  {\bibfnamefont {A.}~\bibnamefont {Liang}}, \bibinfo {author} {\bibfnamefont
  {C.}~\bibnamefont {Chen}},  \emph {et~al.},\ }\href@noop {} {\bibfield
  {journal} {\bibinfo  {journal} {Physical Review X}\ }\textbf {\bibinfo
  {volume} {9}},\ \bibinfo {pages} {041040} (\bibinfo {year}
  {2019}{\natexlab{b}})}\BibitemShut {NoStop}%
\bibitem [{\citenamefont {Li}\ \emph {et~al.}(2019{\natexlab{b}})\citenamefont
  {Li}, \citenamefont {Gao}, \citenamefont {Duan}, \citenamefont {Xu},
  \citenamefont {Zhu}, \citenamefont {Tian}, \citenamefont {Gao}, \citenamefont
  {Fan}, \citenamefont {Rao}, \citenamefont {Huang} \emph
  {et~al.}}]{li2019dirac}%
  \BibitemOpen
  \bibfield  {author} {\bibinfo {author} {\bibfnamefont {H.}~\bibnamefont
  {Li}}, \bibinfo {author} {\bibfnamefont {S.-Y.}\ \bibnamefont {Gao}},
  \bibinfo {author} {\bibfnamefont {S.-F.}\ \bibnamefont {Duan}}, \bibinfo
  {author} {\bibfnamefont {Y.-F.}\ \bibnamefont {Xu}}, \bibinfo {author}
  {\bibfnamefont {K.-J.}\ \bibnamefont {Zhu}}, \bibinfo {author} {\bibfnamefont
  {S.-J.}\ \bibnamefont {Tian}}, \bibinfo {author} {\bibfnamefont {J.-C.}\
  \bibnamefont {Gao}}, \bibinfo {author} {\bibfnamefont {W.-H.}\ \bibnamefont
  {Fan}}, \bibinfo {author} {\bibfnamefont {Z.-C.}\ \bibnamefont {Rao}},
  \bibinfo {author} {\bibfnamefont {J.-R.}\ \bibnamefont {Huang}},  \emph
  {et~al.},\ }\href@noop {} {\bibfield  {journal} {\bibinfo  {journal}
  {Physical Review X}\ }\textbf {\bibinfo {volume} {9}},\ \bibinfo {pages}
  {041039} (\bibinfo {year} {2019}{\natexlab{b}})}\BibitemShut {NoStop}%
\bibitem [{\citenamefont {Li}\ \emph {et~al.}(2020)\citenamefont {Li},
  \citenamefont {Yan}, \citenamefont {Pajerowski}, \citenamefont {Gordon},
  \citenamefont {Nedi{\'c}}, \citenamefont {Sizyuk}, \citenamefont {Ke},
  \citenamefont {Orth}, \citenamefont {Vaknin},\ and\ \citenamefont
  {McQueeney}}]{li2020competing}%
  \BibitemOpen
  \bibfield  {author} {\bibinfo {author} {\bibfnamefont {B.}~\bibnamefont
  {Li}}, \bibinfo {author} {\bibfnamefont {J.-Q.}\ \bibnamefont {Yan}},
  \bibinfo {author} {\bibfnamefont {D.~M.}\ \bibnamefont {Pajerowski}},
  \bibinfo {author} {\bibfnamefont {E.}~\bibnamefont {Gordon}}, \bibinfo
  {author} {\bibfnamefont {A.-M.}\ \bibnamefont {Nedi{\'c}}}, \bibinfo {author}
  {\bibfnamefont {Y.}~\bibnamefont {Sizyuk}}, \bibinfo {author} {\bibfnamefont
  {L.}~\bibnamefont {Ke}}, \bibinfo {author} {\bibfnamefont {P.~P.}\
  \bibnamefont {Orth}}, \bibinfo {author} {\bibfnamefont {D.}~\bibnamefont
  {Vaknin}}, \ and\ \bibinfo {author} {\bibfnamefont {R.~J.}\ \bibnamefont
  {McQueeney}},\ }\href@noop {} {\bibfield  {journal} {\bibinfo  {journal}
  {Physical review letters}\ }\textbf {\bibinfo {volume} {124}},\ \bibinfo
  {pages} {167204} (\bibinfo {year} {2020})}\BibitemShut {NoStop}%
\bibitem [{\citenamefont {Ding}\ \emph {et~al.}(2020)\citenamefont {Ding},
  \citenamefont {Hu}, \citenamefont {Ye}, \citenamefont {Feng}, \citenamefont
  {Ni},\ and\ \citenamefont {Cao}}]{ding2020crystal}%
  \BibitemOpen
  \bibfield  {author} {\bibinfo {author} {\bibfnamefont {L.}~\bibnamefont
  {Ding}}, \bibinfo {author} {\bibfnamefont {C.}~\bibnamefont {Hu}}, \bibinfo
  {author} {\bibfnamefont {F.}~\bibnamefont {Ye}}, \bibinfo {author}
  {\bibfnamefont {E.}~\bibnamefont {Feng}}, \bibinfo {author} {\bibfnamefont
  {N.}~\bibnamefont {Ni}}, \ and\ \bibinfo {author} {\bibfnamefont
  {H.}~\bibnamefont {Cao}},\ }\href@noop {} {\bibfield  {journal} {\bibinfo
  {journal} {Physical Review B}\ }\textbf {\bibinfo {volume} {101}},\ \bibinfo
  {pages} {020412} (\bibinfo {year} {2020})}\BibitemShut {NoStop}%
\bibitem [{\citenamefont {Shi}\ \emph {et~al.}(2019)\citenamefont {Shi},
  \citenamefont {Lei}, \citenamefont {Zhu}, \citenamefont {Ma}, \citenamefont
  {Cui}, \citenamefont {Sun}, \citenamefont {Ying},\ and\ \citenamefont
  {Chen}}]{shi2019magnetic}%
  \BibitemOpen
  \bibfield  {author} {\bibinfo {author} {\bibfnamefont {M.}~\bibnamefont
  {Shi}}, \bibinfo {author} {\bibfnamefont {B.}~\bibnamefont {Lei}}, \bibinfo
  {author} {\bibfnamefont {C.}~\bibnamefont {Zhu}}, \bibinfo {author}
  {\bibfnamefont {D.}~\bibnamefont {Ma}}, \bibinfo {author} {\bibfnamefont
  {J.}~\bibnamefont {Cui}}, \bibinfo {author} {\bibfnamefont {Z.}~\bibnamefont
  {Sun}}, \bibinfo {author} {\bibfnamefont {J.}~\bibnamefont {Ying}}, \ and\
  \bibinfo {author} {\bibfnamefont {X.}~\bibnamefont {Chen}},\ }\href@noop {}
  {\bibfield  {journal} {\bibinfo  {journal} {Physical Review B}\ }\textbf
  {\bibinfo {volume} {100}},\ \bibinfo {pages} {155144} (\bibinfo {year}
  {2019})}\BibitemShut {NoStop}%
\bibitem [{\citenamefont {Tian}\ \emph {et~al.}(2020)\citenamefont {Tian},
  \citenamefont {Gao}, \citenamefont {Nie}, \citenamefont {Qian}, \citenamefont
  {Gong}, \citenamefont {Fu}, \citenamefont {Li}, \citenamefont {Fan},
  \citenamefont {Zhang}, \citenamefont {Kondo}, \citenamefont {Shin},
  \citenamefont {Adell}, \citenamefont {Fedderwitz}, \citenamefont {Ding},
  \citenamefont {Wang}, \citenamefont {Qian},\ and\ \citenamefont
  {Lei}}]{tian2019magnetic}%
  \BibitemOpen
  \bibfield  {author} {\bibinfo {author} {\bibfnamefont {S.}~\bibnamefont
  {Tian}}, \bibinfo {author} {\bibfnamefont {S.}~\bibnamefont {Gao}}, \bibinfo
  {author} {\bibfnamefont {S.}~\bibnamefont {Nie}}, \bibinfo {author}
  {\bibfnamefont {Y.}~\bibnamefont {Qian}}, \bibinfo {author} {\bibfnamefont
  {C.}~\bibnamefont {Gong}}, \bibinfo {author} {\bibfnamefont {Y.}~\bibnamefont
  {Fu}}, \bibinfo {author} {\bibfnamefont {H.}~\bibnamefont {Li}}, \bibinfo
  {author} {\bibfnamefont {W.}~\bibnamefont {Fan}}, \bibinfo {author}
  {\bibfnamefont {P.}~\bibnamefont {Zhang}}, \bibinfo {author} {\bibfnamefont
  {T.}~\bibnamefont {Kondo}}, \bibinfo {author} {\bibfnamefont
  {S.}~\bibnamefont {Shin}}, \bibinfo {author} {\bibfnamefont {J.}~\bibnamefont
  {Adell}}, \bibinfo {author} {\bibfnamefont {H.}~\bibnamefont {Fedderwitz}},
  \bibinfo {author} {\bibfnamefont {H.}~\bibnamefont {Ding}}, \bibinfo {author}
  {\bibfnamefont {Z.}~\bibnamefont {Wang}}, \bibinfo {author} {\bibfnamefont
  {T.}~\bibnamefont {Qian}}, \ and\ \bibinfo {author} {\bibfnamefont
  {H.}~\bibnamefont {Lei}},\ }\href {\doibase 10.1103/PhysRevB.102.035144}
  {\bibfield  {journal} {\bibinfo  {journal} {Phys. Rev. B}\ }\textbf {\bibinfo
  {volume} {102}},\ \bibinfo {pages} {035144} (\bibinfo {year}
  {2020})}\BibitemShut {NoStop}%
\bibitem [{\citenamefont {Yan}\ \emph {et~al.}(2020)\citenamefont {Yan},
  \citenamefont {Liu}, \citenamefont {Parker}, \citenamefont {Wu},
  \citenamefont {Aczel}, \citenamefont {Matsuda}, \citenamefont {McGuire},\
  and\ \citenamefont {Sales}}]{yan2020type}%
  \BibitemOpen
  \bibfield  {author} {\bibinfo {author} {\bibfnamefont {J.-Q.}\ \bibnamefont
  {Yan}}, \bibinfo {author} {\bibfnamefont {Y.}~\bibnamefont {Liu}}, \bibinfo
  {author} {\bibfnamefont {D.}~\bibnamefont {Parker}}, \bibinfo {author}
  {\bibfnamefont {Y.}~\bibnamefont {Wu}}, \bibinfo {author} {\bibfnamefont
  {A.}~\bibnamefont {Aczel}}, \bibinfo {author} {\bibfnamefont
  {M.}~\bibnamefont {Matsuda}}, \bibinfo {author} {\bibfnamefont
  {M.}~\bibnamefont {McGuire}}, \ and\ \bibinfo {author} {\bibfnamefont
  {B.}~\bibnamefont {Sales}},\ }\href@noop {} {\bibfield  {journal} {\bibinfo
  {journal} {Physical Review Materials}\ }\textbf {\bibinfo {volume} {4}},\
  \bibinfo {pages} {054202} (\bibinfo {year} {2020})}\BibitemShut {NoStop}%
\bibitem [{\citenamefont {Gordon}\ \emph {et~al.}(2019)\citenamefont {Gordon},
  \citenamefont {Sun}, \citenamefont {Hu}, \citenamefont {Linn}, \citenamefont
  {Li}, \citenamefont {Liu}, \citenamefont {Liu}, \citenamefont {Mackey},
  \citenamefont {Liu}, \citenamefont {Ni} \emph {et~al.}}]{gordon2019strongly}%
  \BibitemOpen
  \bibfield  {author} {\bibinfo {author} {\bibfnamefont {K.~N.}\ \bibnamefont
  {Gordon}}, \bibinfo {author} {\bibfnamefont {H.}~\bibnamefont {Sun}},
  \bibinfo {author} {\bibfnamefont {C.}~\bibnamefont {Hu}}, \bibinfo {author}
  {\bibfnamefont {A.~G.}\ \bibnamefont {Linn}}, \bibinfo {author}
  {\bibfnamefont {H.}~\bibnamefont {Li}}, \bibinfo {author} {\bibfnamefont
  {Y.}~\bibnamefont {Liu}}, \bibinfo {author} {\bibfnamefont {P.}~\bibnamefont
  {Liu}}, \bibinfo {author} {\bibfnamefont {S.}~\bibnamefont {Mackey}},
  \bibinfo {author} {\bibfnamefont {Q.}~\bibnamefont {Liu}}, \bibinfo {author}
  {\bibfnamefont {N.}~\bibnamefont {Ni}},  \emph {et~al.},\ }\href@noop {}
  {\bibfield  {journal} {\bibinfo  {journal} {arXiv preprint arXiv:1910.13943}\
  } (\bibinfo {year} {2019})}\BibitemShut {NoStop}%
\bibitem [{\citenamefont {Hu}\ \emph {et~al.}(2020{\natexlab{c}})\citenamefont
  {Hu}, \citenamefont {Xu}, \citenamefont {Shi}, \citenamefont {Luo},
  \citenamefont {Peng}, \citenamefont {Wang}, \citenamefont {Ying},
  \citenamefont {Wu}, \citenamefont {Liu}, \citenamefont {Zhang} \emph
  {et~al.}}]{hu2020universal}%
  \BibitemOpen
  \bibfield  {author} {\bibinfo {author} {\bibfnamefont {Y.}~\bibnamefont
  {Hu}}, \bibinfo {author} {\bibfnamefont {L.}~\bibnamefont {Xu}}, \bibinfo
  {author} {\bibfnamefont {M.}~\bibnamefont {Shi}}, \bibinfo {author}
  {\bibfnamefont {A.}~\bibnamefont {Luo}}, \bibinfo {author} {\bibfnamefont
  {S.}~\bibnamefont {Peng}}, \bibinfo {author} {\bibfnamefont {Z.}~\bibnamefont
  {Wang}}, \bibinfo {author} {\bibfnamefont {J.}~\bibnamefont {Ying}}, \bibinfo
  {author} {\bibfnamefont {T.}~\bibnamefont {Wu}}, \bibinfo {author}
  {\bibfnamefont {Z.}~\bibnamefont {Liu}}, \bibinfo {author} {\bibfnamefont
  {C.}~\bibnamefont {Zhang}},  \emph {et~al.},\ }\href@noop {} {\bibfield
  {journal} {\bibinfo  {journal} {Physical Review B}\ }\textbf {\bibinfo
  {volume} {101}},\ \bibinfo {pages} {161113} (\bibinfo {year}
  {2020}{\natexlab{c}})}\BibitemShut {NoStop}%
\bibitem [{\citenamefont {Xu}\ \emph {et~al.}(2019)\citenamefont {Xu},
  \citenamefont {Mao}, \citenamefont {Wang}, \citenamefont {Li}, \citenamefont
  {Chen}, \citenamefont {Xia}, \citenamefont {Li}, \citenamefont {Zhang},
  \citenamefont {Zheng}, \citenamefont {Huang} \emph
  {et~al.}}]{xu2019persistent}%
  \BibitemOpen
  \bibfield  {author} {\bibinfo {author} {\bibfnamefont {L.}~\bibnamefont
  {Xu}}, \bibinfo {author} {\bibfnamefont {Y.}~\bibnamefont {Mao}}, \bibinfo
  {author} {\bibfnamefont {H.}~\bibnamefont {Wang}}, \bibinfo {author}
  {\bibfnamefont {J.}~\bibnamefont {Li}}, \bibinfo {author} {\bibfnamefont
  {Y.}~\bibnamefont {Chen}}, \bibinfo {author} {\bibfnamefont {Y.}~\bibnamefont
  {Xia}}, \bibinfo {author} {\bibfnamefont {Y.}~\bibnamefont {Li}}, \bibinfo
  {author} {\bibfnamefont {J.}~\bibnamefont {Zhang}}, \bibinfo {author}
  {\bibfnamefont {H.}~\bibnamefont {Zheng}}, \bibinfo {author} {\bibfnamefont
  {K.}~\bibnamefont {Huang}},  \emph {et~al.},\ }\href@noop {} {\bibfield
  {journal} {\bibinfo  {journal} {arXiv preprint arXiv:1910.11014}\ } (\bibinfo
  {year} {2019})}\BibitemShut {NoStop}%
\bibitem [{\citenamefont {Jo}\ \emph {et~al.}(2020)\citenamefont {Jo},
  \citenamefont {Wang}, \citenamefont {Slager}, \citenamefont {Yan},
  \citenamefont {Wu}, \citenamefont {Lee}, \citenamefont {Schrunk},
  \citenamefont {Vishwanath},\ and\ \citenamefont
  {Kaminski}}]{jo2020intrinsic}%
  \BibitemOpen
  \bibfield  {author} {\bibinfo {author} {\bibfnamefont {N.~H.}\ \bibnamefont
  {Jo}}, \bibinfo {author} {\bibfnamefont {L.-L.}\ \bibnamefont {Wang}},
  \bibinfo {author} {\bibfnamefont {R.-J.}\ \bibnamefont {Slager}}, \bibinfo
  {author} {\bibfnamefont {J.}~\bibnamefont {Yan}}, \bibinfo {author}
  {\bibfnamefont {Y.}~\bibnamefont {Wu}}, \bibinfo {author} {\bibfnamefont
  {K.}~\bibnamefont {Lee}}, \bibinfo {author} {\bibfnamefont {B.}~\bibnamefont
  {Schrunk}}, \bibinfo {author} {\bibfnamefont {A.}~\bibnamefont {Vishwanath}},
  \ and\ \bibinfo {author} {\bibfnamefont {A.}~\bibnamefont {Kaminski}},\
  }\href@noop {} {\bibfield  {journal} {\bibinfo  {journal} {Physical Review
  B}\ }\textbf {\bibinfo {volume} {102}},\ \bibinfo {pages} {045130} (\bibinfo
  {year} {2020})}\BibitemShut {NoStop}%
\bibitem [{\citenamefont {Klimovskikh}\ \emph {et~al.}(2020)\citenamefont
  {Klimovskikh}, \citenamefont {Otrokov}, \citenamefont {Estyunin},
  \citenamefont {Eremeev}, \citenamefont {Filnov}, \citenamefont {Koroleva},
  \citenamefont {Shevchenko}, \citenamefont {Voroshnin}, \citenamefont
  {Rybkin}, \citenamefont {Rusinov} \emph {et~al.}}]{klimovskikh2020tunable}%
  \BibitemOpen
  \bibfield  {author} {\bibinfo {author} {\bibfnamefont {I.~I.}\ \bibnamefont
  {Klimovskikh}}, \bibinfo {author} {\bibfnamefont {M.~M.}\ \bibnamefont
  {Otrokov}}, \bibinfo {author} {\bibfnamefont {D.}~\bibnamefont {Estyunin}},
  \bibinfo {author} {\bibfnamefont {S.~V.}\ \bibnamefont {Eremeev}}, \bibinfo
  {author} {\bibfnamefont {S.~O.}\ \bibnamefont {Filnov}}, \bibinfo {author}
  {\bibfnamefont {A.}~\bibnamefont {Koroleva}}, \bibinfo {author}
  {\bibfnamefont {E.}~\bibnamefont {Shevchenko}}, \bibinfo {author}
  {\bibfnamefont {V.}~\bibnamefont {Voroshnin}}, \bibinfo {author}
  {\bibfnamefont {A.~G.}\ \bibnamefont {Rybkin}}, \bibinfo {author}
  {\bibfnamefont {I.~P.}\ \bibnamefont {Rusinov}},  \emph {et~al.},\
  }\href@noop {} {\bibfield  {journal} {\bibinfo  {journal} {npj Quantum
  Materials}\ }\textbf {\bibinfo {volume} {5}},\ \bibinfo {pages} {1} (\bibinfo
  {year} {2020})}\BibitemShut {NoStop}%
\bibitem [{\citenamefont {Yan}\ \emph {et~al.}(2019{\natexlab{b}})\citenamefont
  {Yan}, \citenamefont {Okamoto}, \citenamefont {McGuire}, \citenamefont {May},
  \citenamefont {McQueeney},\ and\ \citenamefont {Sales}}]{yan2019evolution}%
  \BibitemOpen
  \bibfield  {author} {\bibinfo {author} {\bibfnamefont {J.-Q.}\ \bibnamefont
  {Yan}}, \bibinfo {author} {\bibfnamefont {S.}~\bibnamefont {Okamoto}},
  \bibinfo {author} {\bibfnamefont {M.~A.}\ \bibnamefont {McGuire}}, \bibinfo
  {author} {\bibfnamefont {A.~F.}\ \bibnamefont {May}}, \bibinfo {author}
  {\bibfnamefont {R.~J.}\ \bibnamefont {McQueeney}}, \ and\ \bibinfo {author}
  {\bibfnamefont {B.~C.}\ \bibnamefont {Sales}},\ }\href@noop {} {\bibfield
  {journal} {\bibinfo  {journal} {Physical Review B}\ }\textbf {\bibinfo
  {volume} {100}},\ \bibinfo {pages} {104409} (\bibinfo {year}
  {2019}{\natexlab{b}})}\BibitemShut {NoStop}%
\bibitem [{\citenamefont {Ko}\ \emph {et~al.}(2020)\citenamefont {Ko},
  \citenamefont {Kolmer}, \citenamefont {Yan}, \citenamefont {Pham},
  \citenamefont {Fu}, \citenamefont {L{\"u}pke}, \citenamefont {Okamoto},
  \citenamefont {Gai}, \citenamefont {Ganesh},\ and\ \citenamefont
  {Li}}]{ko2020realizing}%
  \BibitemOpen
  \bibfield  {author} {\bibinfo {author} {\bibfnamefont {W.}~\bibnamefont
  {Ko}}, \bibinfo {author} {\bibfnamefont {M.}~\bibnamefont {Kolmer}}, \bibinfo
  {author} {\bibfnamefont {J.}~\bibnamefont {Yan}}, \bibinfo {author}
  {\bibfnamefont {A.~D.}\ \bibnamefont {Pham}}, \bibinfo {author}
  {\bibfnamefont {M.}~\bibnamefont {Fu}}, \bibinfo {author} {\bibfnamefont
  {F.}~\bibnamefont {L{\"u}pke}}, \bibinfo {author} {\bibfnamefont
  {S.}~\bibnamefont {Okamoto}}, \bibinfo {author} {\bibfnamefont
  {Z.}~\bibnamefont {Gai}}, \bibinfo {author} {\bibfnamefont {P.}~\bibnamefont
  {Ganesh}}, \ and\ \bibinfo {author} {\bibfnamefont {A.-P.}\ \bibnamefont
  {Li}},\ }\href@noop {} {\bibfield  {journal} {\bibinfo  {journal} {Physical
  Review B}\ }\textbf {\bibinfo {volume} {102}},\ \bibinfo {pages} {115402}
  (\bibinfo {year} {2020})}\BibitemShut {NoStop}%
\bibitem [{\citenamefont {Murakami}\ \emph {et~al.}(2019)\citenamefont
  {Murakami}, \citenamefont {Nambu}, \citenamefont {Koretsune}, \citenamefont
  {Xiangyu}, \citenamefont {Yamamoto}, \citenamefont {Brown},\ and\
  \citenamefont {Kageyama}}]{murakami2019realization}%
  \BibitemOpen
  \bibfield  {author} {\bibinfo {author} {\bibfnamefont {T.}~\bibnamefont
  {Murakami}}, \bibinfo {author} {\bibfnamefont {Y.}~\bibnamefont {Nambu}},
  \bibinfo {author} {\bibfnamefont {T.}~\bibnamefont {Koretsune}}, \bibinfo
  {author} {\bibfnamefont {G.}~\bibnamefont {Xiangyu}}, \bibinfo {author}
  {\bibfnamefont {T.}~\bibnamefont {Yamamoto}}, \bibinfo {author}
  {\bibfnamefont {C.~M.}\ \bibnamefont {Brown}}, \ and\ \bibinfo {author}
  {\bibfnamefont {H.}~\bibnamefont {Kageyama}},\ }\href@noop {} {\bibfield
  {journal} {\bibinfo  {journal} {Physical Review B}\ }\textbf {\bibinfo
  {volume} {100}},\ \bibinfo {pages} {195103} (\bibinfo {year}
  {2019})}\BibitemShut {NoStop}%
\bibitem [{\citenamefont {Liu}\ \emph {et~al.}(2021)\citenamefont {Liu},
  \citenamefont {Wang}, \citenamefont {Zheng}, \citenamefont {Huang},
  \citenamefont {Wang}, \citenamefont {Chi}, \citenamefont {Wu}, \citenamefont
  {Chakoumakos}, \citenamefont {McGuire}, \citenamefont {Sales} \emph
  {et~al.}}]{liu2021site}%
  \BibitemOpen
  \bibfield  {author} {\bibinfo {author} {\bibfnamefont {Y.}~\bibnamefont
  {Liu}}, \bibinfo {author} {\bibfnamefont {L.-L.}\ \bibnamefont {Wang}},
  \bibinfo {author} {\bibfnamefont {Q.}~\bibnamefont {Zheng}}, \bibinfo
  {author} {\bibfnamefont {Z.}~\bibnamefont {Huang}}, \bibinfo {author}
  {\bibfnamefont {X.}~\bibnamefont {Wang}}, \bibinfo {author} {\bibfnamefont
  {M.}~\bibnamefont {Chi}}, \bibinfo {author} {\bibfnamefont {Y.}~\bibnamefont
  {Wu}}, \bibinfo {author} {\bibfnamefont {B.~C.}\ \bibnamefont {Chakoumakos}},
  \bibinfo {author} {\bibfnamefont {M.~A.}\ \bibnamefont {McGuire}}, \bibinfo
  {author} {\bibfnamefont {B.~C.}\ \bibnamefont {Sales}},  \emph {et~al.},\
  }\href@noop {} {\bibfield  {journal} {\bibinfo  {journal} {Physical Review
  X}\ }\textbf {\bibinfo {volume} {11}},\ \bibinfo {pages} {021033} (\bibinfo
  {year} {2021})}\BibitemShut {NoStop}%
\bibitem [{\citenamefont {Du}\ \emph {et~al.}(2021)\citenamefont {Du},
  \citenamefont {Yan}, \citenamefont {Cooper},\ and\ \citenamefont
  {Eisenbach}}]{du2021tuning}%
  \BibitemOpen
  \bibfield  {author} {\bibinfo {author} {\bibfnamefont {M.-H.}\ \bibnamefont
  {Du}}, \bibinfo {author} {\bibfnamefont {J.}~\bibnamefont {Yan}}, \bibinfo
  {author} {\bibfnamefont {V.~R.}\ \bibnamefont {Cooper}}, \ and\ \bibinfo
  {author} {\bibfnamefont {M.}~\bibnamefont {Eisenbach}},\ }\href@noop {}
  {\bibfield  {journal} {\bibinfo  {journal} {Advanced Functional Materials}\
  }\textbf {\bibinfo {volume} {31}},\ \bibinfo {pages} {2006516} (\bibinfo
  {year} {2021})}\BibitemShut {NoStop}%
\bibitem [{\citenamefont {Wu}\ \emph {et~al.}(2020)\citenamefont {Wu},
  \citenamefont {Liu}, \citenamefont {Liu}, \citenamefont {Wang}, \citenamefont
  {Li}, \citenamefont {Lu}, \citenamefont {Matsuishi},\ and\ \citenamefont
  {Hosono}}]{wu2020toward}%
  \BibitemOpen
  \bibfield  {author} {\bibinfo {author} {\bibfnamefont {J.}~\bibnamefont
  {Wu}}, \bibinfo {author} {\bibfnamefont {F.}~\bibnamefont {Liu}}, \bibinfo
  {author} {\bibfnamefont {C.}~\bibnamefont {Liu}}, \bibinfo {author}
  {\bibfnamefont {Y.}~\bibnamefont {Wang}}, \bibinfo {author} {\bibfnamefont
  {C.}~\bibnamefont {Li}}, \bibinfo {author} {\bibfnamefont {Y.}~\bibnamefont
  {Lu}}, \bibinfo {author} {\bibfnamefont {S.}~\bibnamefont {Matsuishi}}, \
  and\ \bibinfo {author} {\bibfnamefont {H.}~\bibnamefont {Hosono}},\
  }\href@noop {} {\bibfield  {journal} {\bibinfo  {journal} {Advanced
  Materials}\ ,\ \bibinfo {pages} {2001815}} (\bibinfo {year}
  {2020})}\BibitemShut {NoStop}%
\bibitem [{\citenamefont {Chakoumakos}\ \emph {et~al.}(2011)\citenamefont
  {Chakoumakos}, \citenamefont {Cao}, \citenamefont {Ye}, \citenamefont
  {Stoica}, \citenamefont {Popovici}, \citenamefont {Sundaram}, \citenamefont
  {Zhou}, \citenamefont {Hicks}, \citenamefont {Lynn},\ and\ \citenamefont
  {Riedel}}]{chakoumakos2011four}%
  \BibitemOpen
  \bibfield  {author} {\bibinfo {author} {\bibfnamefont {B.~C.}\ \bibnamefont
  {Chakoumakos}}, \bibinfo {author} {\bibfnamefont {H.}~\bibnamefont {Cao}},
  \bibinfo {author} {\bibfnamefont {F.}~\bibnamefont {Ye}}, \bibinfo {author}
  {\bibfnamefont {A.~D.}\ \bibnamefont {Stoica}}, \bibinfo {author}
  {\bibfnamefont {M.}~\bibnamefont {Popovici}}, \bibinfo {author}
  {\bibfnamefont {M.}~\bibnamefont {Sundaram}}, \bibinfo {author}
  {\bibfnamefont {W.}~\bibnamefont {Zhou}}, \bibinfo {author} {\bibfnamefont
  {J.~S.}\ \bibnamefont {Hicks}}, \bibinfo {author} {\bibfnamefont {G.~W.}\
  \bibnamefont {Lynn}}, \ and\ \bibinfo {author} {\bibfnamefont {R.~A.}\
  \bibnamefont {Riedel}},\ }\href@noop {} {\bibfield  {journal} {\bibinfo
  {journal} {Journal of Applied Crystallography}\ }\textbf {\bibinfo {volume}
  {44}},\ \bibinfo {pages} {655} (\bibinfo {year} {2011})}\BibitemShut
  {NoStop}%
\bibitem [{\citenamefont
  {Rodr{\'\i}guez-Carvajal}(1993)}]{rodriguez1993recent}%
  \BibitemOpen
  \bibfield  {author} {\bibinfo {author} {\bibfnamefont {J.}~\bibnamefont
  {Rodr{\'\i}guez-Carvajal}},\ }\href@noop {} {\bibfield  {journal} {\bibinfo
  {journal} {Physica B: Condensed Matter}\ }\textbf {\bibinfo {volume} {192}},\
  \bibinfo {pages} {55} (\bibinfo {year} {1993})}\BibitemShut {NoStop}%
\bibitem [{\citenamefont {Bl{\"o}chl}(1994)}]{blochl1994projector}%
  \BibitemOpen
  \bibfield  {author} {\bibinfo {author} {\bibfnamefont {P.~E.}\ \bibnamefont
  {Bl{\"o}chl}},\ }\href@noop {} {\bibfield  {journal} {\bibinfo  {journal}
  {Physical review B}\ }\textbf {\bibinfo {volume} {50}},\ \bibinfo {pages}
  {17953} (\bibinfo {year} {1994})}\BibitemShut {NoStop}%
\bibitem [{\citenamefont {Kresse}\ and\ \citenamefont
  {Joubert}(1999)}]{kresse1999ultrasoft}%
  \BibitemOpen
  \bibfield  {author} {\bibinfo {author} {\bibfnamefont {G.}~\bibnamefont
  {Kresse}}\ and\ \bibinfo {author} {\bibfnamefont {D.}~\bibnamefont
  {Joubert}},\ }\href@noop {} {\bibfield  {journal} {\bibinfo  {journal}
  {Physical review b}\ }\textbf {\bibinfo {volume} {59}},\ \bibinfo {pages}
  {1758} (\bibinfo {year} {1999})}\BibitemShut {NoStop}%
\bibitem [{\citenamefont {Kresse}\ and\ \citenamefont
  {Furthm{\"u}ller}(1996)}]{kresse1996efficiency}%
  \BibitemOpen
  \bibfield  {author} {\bibinfo {author} {\bibfnamefont {G.}~\bibnamefont
  {Kresse}}\ and\ \bibinfo {author} {\bibfnamefont {J.}~\bibnamefont
  {Furthm{\"u}ller}},\ }\href@noop {} {\bibfield  {journal} {\bibinfo
  {journal} {Computational materials science}\ }\textbf {\bibinfo {volume}
  {6}},\ \bibinfo {pages} {15} (\bibinfo {year} {1996})}\BibitemShut {NoStop}%
\bibitem [{\citenamefont {Perdew}\ \emph {et~al.}(1996)\citenamefont {Perdew},
  \citenamefont {Burke},\ and\ \citenamefont
  {Ernzerhof}}]{perdew1996generalized}%
  \BibitemOpen
  \bibfield  {author} {\bibinfo {author} {\bibfnamefont {J.~P.}\ \bibnamefont
  {Perdew}}, \bibinfo {author} {\bibfnamefont {K.}~\bibnamefont {Burke}}, \
  and\ \bibinfo {author} {\bibfnamefont {M.}~\bibnamefont {Ernzerhof}},\
  }\href@noop {} {\bibfield  {journal} {\bibinfo  {journal} {Physical review
  letters}\ }\textbf {\bibinfo {volume} {77}},\ \bibinfo {pages} {3865}
  (\bibinfo {year} {1996})}\BibitemShut {NoStop}%
\bibitem [{\citenamefont {Dudarev}\ \emph {et~al.}(1998)\citenamefont
  {Dudarev}, \citenamefont {Botton}, \citenamefont {Savrasov}, \citenamefont
  {Szotek}, \citenamefont {Temmerman},\ and\ \citenamefont
  {Sutton}}]{dudarev1998electronic}%
  \BibitemOpen
  \bibfield  {author} {\bibinfo {author} {\bibfnamefont {S.}~\bibnamefont
  {Dudarev}}, \bibinfo {author} {\bibfnamefont {G.}~\bibnamefont {Botton}},
  \bibinfo {author} {\bibfnamefont {S.~Y.}\ \bibnamefont {Savrasov}}, \bibinfo
  {author} {\bibfnamefont {Z.}~\bibnamefont {Szotek}}, \bibinfo {author}
  {\bibfnamefont {W.}~\bibnamefont {Temmerman}}, \ and\ \bibinfo {author}
  {\bibfnamefont {A.}~\bibnamefont {Sutton}},\ }\href@noop {} {\bibfield
  {journal} {\bibinfo  {journal} {Physica status solidi (a)}\ }\textbf
  {\bibinfo {volume} {166}},\ \bibinfo {pages} {429} (\bibinfo {year}
  {1998})}\BibitemShut {NoStop}%
\bibitem [{\citenamefont {Marzari}\ and\ \citenamefont
  {Vanderbilt}(1997)}]{marzari1997maximally}%
  \BibitemOpen
  \bibfield  {author} {\bibinfo {author} {\bibfnamefont {N.}~\bibnamefont
  {Marzari}}\ and\ \bibinfo {author} {\bibfnamefont {D.}~\bibnamefont
  {Vanderbilt}},\ }\href@noop {} {\bibfield  {journal} {\bibinfo  {journal}
  {Physical review B}\ }\textbf {\bibinfo {volume} {56}},\ \bibinfo {pages}
  {12847} (\bibinfo {year} {1997})}\BibitemShut {NoStop}%
\bibitem [{\citenamefont {Winter}()}]{webelements}%
  \BibitemOpen
  \bibfield  {author} {\bibinfo {author} {\bibfnamefont {M.}~\bibnamefont
  {Winter}},\ }\href {http://www.webelements.com/} {\enquote {\bibinfo {title}
  {Webelements$^tm$},}\ }\BibitemShut {NoStop}%
\bibitem [{\citenamefont {Hu}\ \emph {et~al.}(2021{\natexlab{a}})\citenamefont
  {Hu}, \citenamefont {Tanatar}, \citenamefont {Prozorov},\ and\ \citenamefont
  {Ni}}]{MBTrelaxation}%
  \BibitemOpen
  \bibfield  {author} {\bibinfo {author} {\bibfnamefont {C.}~\bibnamefont
  {Hu}}, \bibinfo {author} {\bibfnamefont {M.~A.}\ \bibnamefont {Tanatar}},
  \bibinfo {author} {\bibfnamefont {R.}~\bibnamefont {Prozorov}}, \ and\
  \bibinfo {author} {\bibfnamefont {N.}~\bibnamefont {Ni}},\ }\href@noop {}
  {\bibfield  {journal} {\bibinfo  {journal} {arXiv preprint arXiv:2106.08969}\
  } (\bibinfo {year} {2021}{\natexlab{a}})}\BibitemShut {NoStop}%
\bibitem [{\citenamefont {Lai}\ \emph {et~al.}(2021{\natexlab{a}})\citenamefont
  {Lai}, \citenamefont {Ke}, \citenamefont {Yan}, \citenamefont {McDonald},\
  and\ \citenamefont {McQueeney}}]{jiaqiang_highfield}%
  \BibitemOpen
  \bibfield  {author} {\bibinfo {author} {\bibfnamefont {Y.}~\bibnamefont
  {Lai}}, \bibinfo {author} {\bibfnamefont {L.}~\bibnamefont {Ke}}, \bibinfo
  {author} {\bibfnamefont {J.}~\bibnamefont {Yan}}, \bibinfo {author}
  {\bibfnamefont {R.~D.}\ \bibnamefont {McDonald}}, \ and\ \bibinfo {author}
  {\bibfnamefont {R.~J.}\ \bibnamefont {McQueeney}},\ }\href@noop {} {\bibfield
   {journal} {\bibinfo  {journal} {arXiv preprint arXiv:2102.05797}\ }
  (\bibinfo {year} {2021}{\natexlab{a}})}\BibitemShut {NoStop}%
\bibitem [{\citenamefont {Hu}\ \emph {et~al.}(2021{\natexlab{b}})\citenamefont
  {Hu}, \citenamefont {Tanatar}, \citenamefont {Prozorov},\ and\ \citenamefont
  {Ni}}]{hu2021unusual}%
  \BibitemOpen
  \bibfield  {author} {\bibinfo {author} {\bibfnamefont {C.}~\bibnamefont
  {Hu}}, \bibinfo {author} {\bibfnamefont {M.~A.}\ \bibnamefont {Tanatar}},
  \bibinfo {author} {\bibfnamefont {R.}~\bibnamefont {Prozorov}}, \ and\
  \bibinfo {author} {\bibfnamefont {N.}~\bibnamefont {Ni}},\ }\href@noop {}
  {\bibfield  {journal} {\bibinfo  {journal} {arXiv preprint arXiv:2106.08969}\
  } (\bibinfo {year} {2021}{\natexlab{b}})}\BibitemShut {NoStop}%
\bibitem [{\citenamefont {Ding}\ \emph {et~al.}(2021)\citenamefont {Ding},
  \citenamefont {Hu}, \citenamefont {Feng}, \citenamefont {Jiang},
  \citenamefont {Kibalin}, \citenamefont {Gukasov}, \citenamefont {Chi},
  \citenamefont {Ni},\ and\ \citenamefont {Cao}}]{ding2021neutron}%
  \BibitemOpen
  \bibfield  {author} {\bibinfo {author} {\bibfnamefont {L.}~\bibnamefont
  {Ding}}, \bibinfo {author} {\bibfnamefont {C.}~\bibnamefont {Hu}}, \bibinfo
  {author} {\bibfnamefont {E.}~\bibnamefont {Feng}}, \bibinfo {author}
  {\bibfnamefont {C.}~\bibnamefont {Jiang}}, \bibinfo {author} {\bibfnamefont
  {I.~A.}\ \bibnamefont {Kibalin}}, \bibinfo {author} {\bibfnamefont
  {A.}~\bibnamefont {Gukasov}}, \bibinfo {author} {\bibfnamefont
  {M.}~\bibnamefont {Chi}}, \bibinfo {author} {\bibfnamefont {N.}~\bibnamefont
  {Ni}}, \ and\ \bibinfo {author} {\bibfnamefont {H.}~\bibnamefont {Cao}},\
  }\href@noop {} {\bibfield  {journal} {\bibinfo  {journal} {Journal of Physics
  D: Applied Physics}\ }\textbf {\bibinfo {volume} {54}},\ \bibinfo {pages}
  {174003} (\bibinfo {year} {2021})}\BibitemShut {NoStop}%
\bibitem [{\citenamefont {Mong}\ \emph {et~al.}(2010)\citenamefont {Mong},
  \citenamefont {Essin},\ and\ \citenamefont
  {Moore}}]{mong2010antiferromagnetic}%
  \BibitemOpen
  \bibfield  {author} {\bibinfo {author} {\bibfnamefont {R.~S.}\ \bibnamefont
  {Mong}}, \bibinfo {author} {\bibfnamefont {A.~M.}\ \bibnamefont {Essin}}, \
  and\ \bibinfo {author} {\bibfnamefont {J.~E.}\ \bibnamefont {Moore}},\
  }\href@noop {} {\bibfield  {journal} {\bibinfo  {journal} {Physical Review
  B}\ }\textbf {\bibinfo {volume} {81}},\ \bibinfo {pages} {245209} (\bibinfo
  {year} {2010})}\BibitemShut {NoStop}%
\bibitem [{\citenamefont {Fang}\ \emph {et~al.}(2013)\citenamefont {Fang},
  \citenamefont {Gilbert},\ and\ \citenamefont
  {Bernevig}}]{fang2013topological}%
  \BibitemOpen
  \bibfield  {author} {\bibinfo {author} {\bibfnamefont {C.}~\bibnamefont
  {Fang}}, \bibinfo {author} {\bibfnamefont {M.~J.}\ \bibnamefont {Gilbert}}, \
  and\ \bibinfo {author} {\bibfnamefont {B.~A.}\ \bibnamefont {Bernevig}},\
  }\href@noop {} {\bibfield  {journal} {\bibinfo  {journal} {Physical Review
  B}\ }\textbf {\bibinfo {volume} {88}},\ \bibinfo {pages} {085406} (\bibinfo
  {year} {2013})}\BibitemShut {NoStop}%
\bibitem [{\citenamefont {Gui}\ \emph {et~al.}(2019)\citenamefont {Gui},
  \citenamefont {Pletikosic}, \citenamefont {Cao}, \citenamefont {Tien},
  \citenamefont {Xu}, \citenamefont {Zhong}, \citenamefont {Wang},
  \citenamefont {Chang}, \citenamefont {Jia}, \citenamefont {Valla} \emph
  {et~al.}}]{gui2019new}%
  \BibitemOpen
  \bibfield  {author} {\bibinfo {author} {\bibfnamefont {X.}~\bibnamefont
  {Gui}}, \bibinfo {author} {\bibfnamefont {I.}~\bibnamefont {Pletikosic}},
  \bibinfo {author} {\bibfnamefont {H.}~\bibnamefont {Cao}}, \bibinfo {author}
  {\bibfnamefont {H.-J.}\ \bibnamefont {Tien}}, \bibinfo {author}
  {\bibfnamefont {X.}~\bibnamefont {Xu}}, \bibinfo {author} {\bibfnamefont
  {R.}~\bibnamefont {Zhong}}, \bibinfo {author} {\bibfnamefont
  {G.}~\bibnamefont {Wang}}, \bibinfo {author} {\bibfnamefont {T.-R.}\
  \bibnamefont {Chang}}, \bibinfo {author} {\bibfnamefont {S.}~\bibnamefont
  {Jia}}, \bibinfo {author} {\bibfnamefont {T.}~\bibnamefont {Valla}},  \emph
  {et~al.},\ }\href@noop {} {\bibfield  {journal} {\bibinfo  {journal} {ACS
  central science}\ }\textbf {\bibinfo {volume} {5}},\ \bibinfo {pages} {900}
  (\bibinfo {year} {2019})}\BibitemShut {NoStop}%
\bibitem [{\citenamefont {Xu}\ \emph {et~al.}(2011)\citenamefont {Xu},
  \citenamefont {Xia}, \citenamefont {Wray}, \citenamefont {Jia}, \citenamefont
  {Meier}, \citenamefont {Dil}, \citenamefont {Osterwalder}, \citenamefont
  {Slomski}, \citenamefont {Bansil}, \citenamefont {Lin} \emph
  {et~al.}}]{xu2011topological}%
  \BibitemOpen
  \bibfield  {author} {\bibinfo {author} {\bibfnamefont {S.-Y.}\ \bibnamefont
  {Xu}}, \bibinfo {author} {\bibfnamefont {Y.}~\bibnamefont {Xia}}, \bibinfo
  {author} {\bibfnamefont {L.}~\bibnamefont {Wray}}, \bibinfo {author}
  {\bibfnamefont {S.}~\bibnamefont {Jia}}, \bibinfo {author} {\bibfnamefont
  {F.}~\bibnamefont {Meier}}, \bibinfo {author} {\bibfnamefont
  {J.}~\bibnamefont {Dil}}, \bibinfo {author} {\bibfnamefont {J.}~\bibnamefont
  {Osterwalder}}, \bibinfo {author} {\bibfnamefont {B.}~\bibnamefont
  {Slomski}}, \bibinfo {author} {\bibfnamefont {A.}~\bibnamefont {Bansil}},
  \bibinfo {author} {\bibfnamefont {H.}~\bibnamefont {Lin}},  \emph {et~al.},\
  }\href@noop {} {\bibfield  {journal} {\bibinfo  {journal} {Science}\ }\textbf
  {\bibinfo {volume} {332}},\ \bibinfo {pages} {560} (\bibinfo {year}
  {2011})}\BibitemShut {NoStop}%
\bibitem [{\citenamefont {Chang}\ \emph {et~al.}(2016)\citenamefont {Chang},
  \citenamefont {Xu}, \citenamefont {Chang}, \citenamefont {Lee}, \citenamefont
  {Huang}, \citenamefont {Wang}, \citenamefont {Bian}, \citenamefont {Zheng},
  \citenamefont {Sanchez}, \citenamefont {Belopolski} \emph
  {et~al.}}]{chang2016prediction}%
  \BibitemOpen
  \bibfield  {author} {\bibinfo {author} {\bibfnamefont {T.-R.}\ \bibnamefont
  {Chang}}, \bibinfo {author} {\bibfnamefont {S.-Y.}\ \bibnamefont {Xu}},
  \bibinfo {author} {\bibfnamefont {G.}~\bibnamefont {Chang}}, \bibinfo
  {author} {\bibfnamefont {C.-C.}\ \bibnamefont {Lee}}, \bibinfo {author}
  {\bibfnamefont {S.-M.}\ \bibnamefont {Huang}}, \bibinfo {author}
  {\bibfnamefont {B.}~\bibnamefont {Wang}}, \bibinfo {author} {\bibfnamefont
  {G.}~\bibnamefont {Bian}}, \bibinfo {author} {\bibfnamefont {H.}~\bibnamefont
  {Zheng}}, \bibinfo {author} {\bibfnamefont {D.~S.}\ \bibnamefont {Sanchez}},
  \bibinfo {author} {\bibfnamefont {I.}~\bibnamefont {Belopolski}},  \emph
  {et~al.},\ }\href@noop {} {\bibfield  {journal} {\bibinfo  {journal} {Nature
  communications}\ }\textbf {\bibinfo {volume} {7}},\ \bibinfo {pages} {1}
  (\bibinfo {year} {2016})}\BibitemShut {NoStop}%
\bibitem [{\citenamefont {Turner}\ \emph {et~al.}(2012)\citenamefont {Turner},
  \citenamefont {Zhang}, \citenamefont {Mong},\ and\ \citenamefont
  {Vishwanath}}]{turner2012quantized}%
  \BibitemOpen
  \bibfield  {author} {\bibinfo {author} {\bibfnamefont {A.~M.}\ \bibnamefont
  {Turner}}, \bibinfo {author} {\bibfnamefont {Y.}~\bibnamefont {Zhang}},
  \bibinfo {author} {\bibfnamefont {R.~S.}\ \bibnamefont {Mong}}, \ and\
  \bibinfo {author} {\bibfnamefont {A.}~\bibnamefont {Vishwanath}},\
  }\href@noop {} {\bibfield  {journal} {\bibinfo  {journal} {Physical Review
  B}\ }\textbf {\bibinfo {volume} {85}},\ \bibinfo {pages} {165120} (\bibinfo
  {year} {2012})}\BibitemShut {NoStop}%
\bibitem [{\citenamefont {Wieder}\ and\ \citenamefont
  {Bernevig}(2018)}]{wieder2018axion}%
  \BibitemOpen
  \bibfield  {author} {\bibinfo {author} {\bibfnamefont {B.~J.}\ \bibnamefont
  {Wieder}}\ and\ \bibinfo {author} {\bibfnamefont {B.~A.}\ \bibnamefont
  {Bernevig}},\ }\href@noop {} {\bibfield  {journal} {\bibinfo  {journal}
  {arXiv preprint arXiv:1810.02373}\ } (\bibinfo {year} {2018})}\BibitemShut
  {NoStop}%
\bibitem [{sup()}]{supplement}%
  \BibitemOpen
  \href@noop {} {\bibinfo  {journal} {See Supplemental Material}\ }\BibitemShut
  {NoStop}%
\bibitem [{\citenamefont {Varnava}\ and\ \citenamefont
  {Vanderbilt}(2018)}]{varnava2018surfaces}%
  \BibitemOpen
\bibfield  {journal} {  }\bibfield  {author} {\bibinfo {author} {\bibfnamefont
  {N.}~\bibnamefont {Varnava}}\ and\ \bibinfo {author} {\bibfnamefont
  {D.}~\bibnamefont {Vanderbilt}},\ }\href@noop {} {\bibfield  {journal}
  {\bibinfo  {journal} {Physical Review B}\ }\textbf {\bibinfo {volume} {98}},\
  \bibinfo {pages} {245117} (\bibinfo {year} {2018})}\BibitemShut {NoStop}%
\bibitem [{\citenamefont {Lai}\ \emph {et~al.}(2021{\natexlab{b}})\citenamefont
  {Lai}, \citenamefont {Ke}, \citenamefont {Yan}, \citenamefont {McDonald},\
  and\ \citenamefont {McQueeney}}]{lai2021defect}%
  \BibitemOpen
  \bibfield  {author} {\bibinfo {author} {\bibfnamefont {Y.}~\bibnamefont
  {Lai}}, \bibinfo {author} {\bibfnamefont {L.}~\bibnamefont {Ke}}, \bibinfo
  {author} {\bibfnamefont {J.}~\bibnamefont {Yan}}, \bibinfo {author}
  {\bibfnamefont {R.~D.}\ \bibnamefont {McDonald}}, \ and\ \bibinfo {author}
  {\bibfnamefont {R.~J.}\ \bibnamefont {McQueeney}},\ }\href@noop {} {\bibfield
   {journal} {\bibinfo  {journal} {arXiv preprint arXiv:2102.05797}\ }
  (\bibinfo {year} {2021}{\natexlab{b}})}\BibitemShut {NoStop}%
\end{thebibliography}%

\end{document}